\documentclass[pdftex,twocolumn,epjc3]{svjour3}  

\RequirePackage[T1]{fontenc}

\usepackage{graphicx}
\RequirePackage{amsmath}
\RequirePackage{mathptmx}      
\RequirePackage{flushend}
\RequirePackage[numbers,sort&compress]{natbib}
\RequirePackage[colorlinks,citecolor=blue,urlcolor=blue,linkcolor=blue]{hyperref}

\usepackage{relsize}
\def\babar{\mbox{\slshape B\kern-0.1em{\smaller A}\kern-0.1em
    B\kern-0.1em{\smaller A\kern-0.2em R}}}

\journalname{Eur. Phys. J. C}
\begin{document}

\title{\boldmath The $\eta$ transition form factor from space- and time-like experimental data}

\author{R.~Escribano\thanksref{e1,addr1} 
        \and 
        P.~Masjuan\thanksref{e2,addr2}
        \and
        P.~Sanchez-Puertas\thanksref{e3,addr2}
}

\thankstext{e1}{e-mail: rescriba@ifae.es}
\thankstext{e2}{e-mail: masjuan@kph.uni-mainz.de}
\thankstext{e3}{e-mail: sanchezp@kph.uni-mainz.de}

\institute{Grup de F\'isica Te\`orica (Departament de F\'isica) and Institut de F\'{\i}sica d'Altes Energies (IFAE),
              Universitat Aut\`onoma de Barcelona, E-08193 Bellaterra (Barcelona), Spain\label{addr1}
                 \and
              PRISMA Cluster of Excellence, Institut f\"ur Kernphysik, Johannes
              Gutenberg-Universit\"at, D-55099 Mainz, Germany \label{addr2}
}    
    
\date{Received: date / Accepted: date}

\maketitle

\begin{abstract}
The $\eta$ transition form factor is analyzed for the first time in both space- and time-like regions
at low and intermediate energies in a model-independent approach through the use of 
rational approximants.
The $\eta\rightarrow e^+e^-\gamma$ experimental data provided by the A2 Collaboration
in the very low-energy region of the dilelectron invariant mass distribution allows for the extraction
of the most precise up-to-date slope and curvature parameters of the form factors as well as 
their values at zero and infinity. 
The impact of these new results on the mixing parameters of the $\eta$-$\eta^\prime$ system,
together with the role played by renormalisation dependent effects,
and on the determination of the $VP\gamma$ couplings from
$V\to P\gamma$ and $P\to V\gamma$ radiative decays are also discussed.

\keywords{
Transition Form Factors, Pad\'e Approximants, 
$\eta$-$\eta^\prime$ Mixing, Radiative Decays.
}

\end{abstract}

\section{Introduction}
\label{Intro}

The pseudoscalar transition form factors (TFFs) describe the effect of the strong interaction
on the $\gamma^*\gamma^*P$ vertex, where $P=\pi^0, \eta, \eta^\prime, \eta_c\ldots$,
and is represented by $F_{P\gamma^*\gamma^*} (q_1^2,q_2^2)$, a function of the photon virtualities $q_1^2$, and $q_2^2$.
From the experimental point of view,
one can study such TFFs from both space- and time-like energy regions.
The time-like region of the TFF can be accessed at meson facilities either through 
the double Dalitz decay processes $P \to l^+l^- l^+l^-$,
which give access to both photon virtualities $(q_1^2,q_2^2)$ in the range
$4m_l^2<(q_1^2,q_2^2)<(m_P-2m_l)^2$,
or the single Dalitz decay processes $P \to l^+l^- \gamma$, 
which contains a single virtual photon with transferred momentum in the range
$4m_l^2<q_1^2<m_P^2$, thus simplifying the TFF to
$F_{P\gamma^\ast\gamma^\ast}(q_1^2,0)\equiv F_{P\gamma^\ast\gamma}(q^2)$.
To complete the time-like region, $e^+e^-$ colliders access to the values $q^2>m_P^2$
through the $e^+e^- \to P\gamma$ annihilation processes.
The space-like region of the TFFs are accessed in  $e^+e^-$ colliders by the two-photon-fusion reaction $e^+e^-\to e^+e^-P$, where at the moment 
the measurement of both virtualities is still an experimental challenge.
The common practice is then to extract the TFF when one of the outgoing leptons is tagged
and the other is not, that is, the single-tag method.
The tagged lepton emits a highly off-shell photon with transferred momentum
$q_1^2\equiv -Q^2$ and is detected,
while the other, untagged, is scattered at a small angle with $q_2^2\simeq 0$.
The form factor extracted from the single-tag experiment is then 
$F_{P\gamma^\ast\gamma^\ast}(-Q^2,0)$ $\equiv F_{P\gamma^\ast\gamma}(Q^2)$.

At low-momentum transfer, the TFF can be described by the expansion
\begin{equation}
\label{eq:tffexpansion}
F_{P\gamma^\ast\gamma}(Q^2)=F_{P\gamma\gamma}(0)
\left(1-b_P\frac{Q^2}{m_P^2}+c_{P}\frac{Q^4}{m_P^4}-d_{P}\frac{Q^6}{m_P^6}+\cdots\right)\ , 
\end{equation}
where $F_{P\gamma\gamma}(0)$ is the normalization,
the low-energy parameters (LEPs) $b_P$, $c_P$ and $d_P$ are 
the slope, the curvature and the third derivative of the TFF, respectively,
and $m_P$ is the pseudoscalar meson mass.
$F_{P\gamma\gamma}(0)$ can be obtained either from the measured two-photon partial width 
of the meson $P$,
\begin{equation}
\label{eq:tffdecay}
|F_{P\gamma\gamma}(0)|^2=
\frac{64 \pi}{(4\pi\alpha)^2}\frac{\Gamma(P\to\gamma\gamma)}{m_P^3}\ ,
\end{equation}
or, in the case of $\pi^0$, $\eta$ and $\eta^\prime$,
from the prediction of the axial anomaly in the chiral limit of QCD.

In this work we shall focus on the $\eta$ TFF exclusively. 
Its slope parameter has been extensively discussed from both theoretical analyses~\cite{
Ametller:1991jv,Czyz:2012nq,Escribano:2013kba,Hanhart:2013vba,Klopot:2013laa}
and experimental measurements~\cite{Dzhelyadin:1980kh,Behrend:1990sr,Gronberg:1997fj,Arnaldi:2009aa,Berghauser:2011zz,
Hodana:2012rc,Aguar-Bartolome:2013vpw}.
On the theory side, 
Chiral Perturbation Theory (ChPT) predicts
$b_{\eta}=0.51$ at $\mu^2=0.69$ GeV$^2$ and for $\sin\theta_P=-1/3$ \cite{Ametller:1991jv}, 
being $\mu$ the renormalisation scale and $\theta_P$ the $\eta$-$\eta^\prime$ mixing angle.
Other theoretical predictions are~\cite{Ametller:1991jv}:
$b_{\eta}=0.53$ from Vector Meson Dominance (VMD),
$b_{\eta}=0.51$ from constituent-quark loops (QL), and
$b_{\eta}=0.36$ from the Brodsky-Lepage (BL) interpolation formula \cite{Brodsky:1981rp}.
Recently, the slope has been predicted to be
$b_{\eta}=0.546(9)$ and $b_{\eta}=0.521(2)$
from a chiral theory with one and two octets of vector resonances \cite{Czyz:2012nq}, respectively,
$b_{\eta}=0.60(6)_{\rm stat}(3)_{\rm sys}$ from rational approximants \cite{Escribano:2013kba},
$b_{\eta}=0.62^{+0.07}_{-0.03}$~\cite{Hanhart:2013vba} and  
$b_{\eta}=0.57^{+0.06}_{-0.03}$~\cite{Kubis:2015sga} from dispersive analyses, and
$b_{\eta}=0.51$ or $0.54$, depending on the data set used as input, from anomaly sum rules
\cite{Klopot:2013laa}.
With respect to the experimental determinations,
the values for the slope are usually obtained after a fit to data using a normalized, 
single-pole term with an associated mass $\Lambda_P$,
\textit{i.e.}
\begin{equation}
\label{eq:tffpole}
F_{P\gamma^*\gamma}(Q^2)=\frac{F_{P\gamma\gamma}(0)}{1+Q^2/\Lambda_P^2}\ .
\end{equation}
The results are
$b_{\eta}=0.428(89)$ from CELLO \cite{Behrend:1990sr} and
$b_{\eta}=0.501(38)$ from CLEO \cite{Gronberg:1997fj}, 
both from space-like data, and
$b_{\eta}=0.57(12)$ from Lepton-G \cite{Dzhelyadin:1980kh}, 
$b_{\eta}=0.585(51)$ from NA60 \cite{Arnaldi:2009aa}
$b_{\eta}=0.58(11)$ from A2 \cite{Berghauser:2011zz}, and
$b_{\eta}=0.68(26)$ from WASA \cite{Hodana:2012rc},
all of them from time-like data.
More recently, the A2 Collaboration reported $b_{\eta}=0.59(5)$ \cite{Aguar-Bartolome:2013vpw},
the most precise experimental determination up to date. 
The curvature was for the first time reported in Ref.~\cite{Escribano:2013kba}
with the value $c_{\eta}=0.37(10)_{\rm stat}(7)_{\rm sys}$.
Nothing is yet reported about the third derivative of the TFF, $d_{\eta}$,
although its role on hadronic quantities where TFFs are important suggests also to look at it 
(see Ref.~\cite{Escribano:2013kba} for its role on the $\eta$ contribution to the 
hadronic light-by-light scattering piece to the muon $(g-2)$).

Several attempts to describe the $\eta$ TFF are available in the literature at present \cite{
Kroll:2010bf,Dorokhov:2011zf,Brodsky:2011yv,Brodsky:2011xx,Klopot:2011qq,Wu:2011gf,
Noguera:2011fv,Balakireva:2011wp,Czyz:2012nq,Melikhov:2012bg,Kroll:2013iwa,
Melikhov:2012qp,Geng:2012qg,Klopot:2012hd,Hanhart:2013vba,Klopot:2013laa,Bijnens:1988kx,Bijnens:1989jb} 
but none of them tries for a unique description of both space- and time-like experimental data,
specially at low energies.
In Ref.~\cite{Masjuan:2012wy}, it was suggested for the $\pi^0$ case that a 
model-independent approach to the space-like TFF can be achieved using a sequence of 
rational functions, the Pad\'e Approximants (PAs), to fit the data.
Later on, in Ref.~\cite{Escribano:2013kba},
the same method was applied to the $\eta$ and $\eta^\prime$ TFFs.
More recently, the A2 Collaboration reported a new measurement of the 
$\eta\to e^+e^-\gamma$ Dalitz decay process 
with the best statistical accuracy up to date \cite{Aguar-Bartolome:2013vpw}.
A comparison with different theoretical approaches was also performed. 
In particular, the results from Ref.~\cite{Escribano:2013kba},
based on space-like data,
were extrapolated to the time-like region and agreed perfectly with their measurement.
Triggered by these new A2 results,
we explore in the present work a combined description of both space- and time-like regions 
of the $\eta$ TFF within our method of rational approximants.
This will provide, for the first time, a determination of the energy dependence of the $\eta$ TFF 
in both regions together with a unified extraction of its LEPs.

Our approach makes use of PAs as fitting functions to all the experimental data.
PAs are rational functions $P^N_M(Q^2)$
(ratio of a polynomial $T_N(Q^2)$ of order $N$ and a polynomial $R_M(Q^2)$ of order $M$)
constructed in such a way that they have the same Taylor expansion as the function
to be approximated up to order ${\cal O}(Q^2)^{N+M+1}$ \cite{Baker}.
Since PAs are built in our case from the unknown low-energy parameters (LEPs) of the TFF, 
once the fit to the experimental data is done,
the reexpansion of the PAs yields the desired coefficients.
Being rational functions the PAs are analytic everywhere except where the poles are located.
Branch cuts cannot in principle be described by PAs, however, if the function to be approximated is of a certain kind, for instance a Stieltjes function, it can be proven mathematically that an infinite order PA is able to reproduce the cut~\cite{Baker}\footnote{
In this case, the infinite number of poles and zeros are seen to be located along the branch cut with the first pole located at the beginning of the branch point.}.
Another interesting issue is the implementation of chiral logarithms of the kind $\log(Q^2/M^2)$, appearing for instance in chiral expansions at next-to-leading order, in the PAs method. These chiral logs admit a Taylor expansion which can be seen as an infinite order diagonal PA and is convergent for any value of $Q^2>0$. Therefore, in case the approximated function includes chiral logs their effects are incorporated in the PAs to a good extent (more precise as the order of the PA increases)\footnote{
For values of $Q^2/M^2\sim 1$ the relative error between the chiral $\log$ and an associated, for instance, second-order PA is of the order of the per mille.}.
The advantage of PAs over Taylor expansions is their ability to enlarge the domain of convergence. However, to prove the convergence of a given PA sequence is a difficult task and only for certain classes of functions this can be done rigorously.
In practice, the success of PAs in the description of experimental data can only be seen \emph{a posteriori} in the sense that the pattern of convergence can be shown but unfortunately not proven mathematically. We refer the interested reader to Refs.~\cite{Masjuan:2008fv,Queralt:2010sv} 
for details on this technique.

In this work, we resume our method \cite{Escribano:2013kba} for fitting the $\eta$ TFF experimental data
after including all the recent available time-like measurements from $\eta\to l^+ l^- \gamma$ decays ($l=e,\mu$).
Besides recapitulating the main features of the method we will address the following issues:
\begin{itemize}
\item
A reevaluation of the systematic errors considered in our previous work is demanded by 
the inclusion of time-like data at these low energies.
This new set of data being more precise than the space-like one, 
its incorporation will allow for an improved systematic error associated with each element of a given PA sequence 
and the increase in order of the sequence itself.
\item
The better description of the low-energy region of the TFF allows for an improved determination of its value at 
zero momentum transfer, which is related to the two-photon decay width of the $\eta$.
The impact of the recent measurement of this width by the KLOE collaboration \cite{Babusci:2012ik}
and of older measurements based on Primakoff techniques is commented.
\item
The role played by high-energy space-like data in view of the fact that in such region only \babar \  data is available.
Related to this, the existing puzzle between the precise mixing scenario derived from the TFF
in contrast to the measured time-like cross section by \babar \  at $q^2=112$~GeV$^2$ \cite{Aubert:2006cy} is discussed.
The possibility for the Belle Collaboration to measure the time-like $\eta$ TFF is also mentioned;
\item
The extraction of the $\eta$-$\eta^\prime$ mixing parameters from the TFFs and the two-photon decays
after discussing the role of the renormalization scale dependence of the singlet decay constant $F_0$.
The new results are much better constrained with the inclusion of time-like experimental data
and turn out to be competitive with standard determinations, such as for instance the analysis of 
$V\to P\gamma$ $(V=\rho, \omega, \phi)$ and $P\to V\gamma$ decays \cite{Escribano:2005qq}.
\item
The determination of these $VP\gamma$ coupling constants from the former mixing parameters
and its comparison with current experimental values.
The effect of OZI-violating parameters and higher-order effects is also discussed.
\end{itemize}

The paper is organized as follows.
In Sect.~\ref{sec:syserror},
a reanalysis of the systematic error related to our method when taking into account
both space- and time-like experimental data is performed.
In Sect.~\ref{sec:PSTFF},
a brief description of the general method for extracting the low-energy parameters of the $\eta$ TFF
using rational approximants is presented and then the impact on them of both
the $\eta\to\gamma\gamma$ latest measurement and the high-energy space-like data is discussed.
In Sect.~\ref{sec:mixing},
the implications of our new results for the determination of the $\eta$-$\eta^\prime$ mixing parameters,
the understanding of the \babar \  puzzle, and the prediction of the $VP\gamma$ couplings are examined.
Finally, in Sect.~\ref{sec:conc} the conclusions of the present analysis are given.

\section{A new systematic error}
\label{sec:syserror}

In the context of Pad\'e approximants, by \emph{systematic error}
it is meant the difference between the function to be approximated and the highest approximant reached after the fit procedure.
If there is seen convergence, the larger the PA order, the smaller the systematic error.
Therefore, any finite order PA should have a definite systematic error.
In this section, we discuss how to obtain such an error for a scenario containing both time- and space-like data.

In order to illustrate the utility of the PA as fitting functions, Ref.~\cite{Masjuan:2012wy}
simulates the real situation of the experimental data on the space-like region by generating with 
different models a set of pseudodata.
Such data were then fitted with a $P^L_1(Q^2)$ (single-pole approximants) sequence 
and the LEPs where extracted. 
This exercise was twofold: 
first it was meant to show the ability of the PA sequence to extract the LEPs and, second,
also provided a systematic error for the extraction of each LEP at each value of $L$.
In Ref.~\cite{Escribano:2013kba}, 
more examples were worked out and further discussed and we refer the interested reader to such references.

Dealing now with a larger set of data, such systematic errors should be reanalyzed, specially because
the amount of time-like data, which covers the lowest energy region
---and is most important for LEPs extraction---
is larger than the space-like one.

Following the strategy presented in Refs.~\cite{Masjuan:2012wy,Escribano:2013kba}, 
we simulate with an holographic model (see~\ref{AppConv} for details
on the model together with on the simulation) the situation of the experimental data from both
space-~\cite{Behrend:1990sr,Gronberg:1997fj,BABAR:2011ad} and 
time-like~\cite{Arnaldi:2009aa,Berghauser:2011zz,Aguar-Bartolome:2013vpw} data.
The results obtained with the holographic model described in~\ref{AppConv}
are collected in Table~\ref{tab:syserror} 
where the relative errors for the first three derivatives for each element on the $P^L_1(Q^2)$ sequence
are reported.
These results are model dependent.
Using, instead, the quark model considered in Ref.~\cite{Masjuan:2012wy},
we find faster convergence and we reach systematic errors one order of magnitude better 
for the higher PA of the sequence than the holographic model.
We chose to use the results with the holographic one to be on the conservative side.

The strategy is then to generate pseudodata for both regions trying to emulate the real experimental situation.
In the space-like region,
we evaluated the model at 10 points in the region $0.6\le Q^2 \le 2.2$ GeV$^2$,
15 points in the region $2.7\le Q^2 \le 7.5$ GeV$^2$, and
9 more points in the region $9\le Q^2 \le 34$ GeV$^2$.
In the time-like region,
the model is evaluated at 8 points in the region $(0.045)^2\le Q^2 \le (0.100)^2$ GeV$^2$,
15 points in the region $(0.115)^2\le Q^2 \le (0.220)^2$ GeV$^2$, and
31 more points in the region $(0.230)^2\le Q^2 \le (0.470)^2$ GeV$^2$.
On top of these set of data points we add the value of $F_{\eta\gamma\gamma}(0,0)$.
All these data points have zero error because we want to obtain a pure systematic error on our fitting functions.
Notice that the majority of points lie in the low-energy region.
This simple exercise also prevents us against over-fitting problems.
The very same study can be performed to evaluate the $P^N_N(Q^2)$ sequence. The results are, however, an order of magnitude better than for the $P^L_1(Q^2)$ one (see the comparison in ~\ref{AppConv}). From now on, we consider only the systematic errors from the latter to be on the conservative side.

The forthcoming BESIII data on the $\eta$ TFF at space-like region below $9$ GeV$^2$ might demand a
reanalysis of our systematic errors, although we think that including them would not really modify our percentages
beyond the precision we are reporting them in Table~\ref{tab:syserror}.
Their data will be, nevertheless, crucial to reduce our statistical errors which is by now the dominant source.

In passing, we also study what would be the systematic error done by a VMD fit to only the time-like data set.
From the three models considered in Refs.~\cite{Masjuan:2012wy,Escribano:2013kba},
the most conservative systematic error found is around $5\%$
(details are presented in~\ref{AppConv}).
Notice that  when fitting space-like data with a VMD such error is around $40\%$.
The reason of such difference is simple because available time-like data is much closer to the origin of energies
than the space-like one and less sensible to higher-order effects.

\begin{table}
\begin{center}
\begin{tabular}{|c||c|c|c|c|c|c|c|c|c|}
\hline
L & $1$ & $2$ & $3$ & $4$ & $5$ & $6$ & $7$ & $8$ & $9$ \\
\hline
$b_{\eta}$ & $9.6$ & $7.0$ & $4.3$ & $3.0$ & $1.8$ & $1.1$ & $0.7$ & $0.4$ & $0.2$ \\
$c_{\eta}$ & $-$ & $4.0$ & $4.0$ & $3.5$ & $2.7$ & $2.0$ & $1.4$ & $0.8$ & $0.5$ \\
$d_{\eta}$ & $-$ & $-$ & $22.2$ & $18.9$ & $14.6$ & $11.3$ & $8.6$ & $5.9$ & $4.0$ \\
\hline
\end{tabular}
\end{center}
\caption{Collection of systematic errors (in percentage $\%$) of the first three derivatives
$b_{\eta},c_{\eta}$ and $d_{\eta}$ of the
$Q^2 F_{\eta \gamma^* \gamma}(Q^2)$ for a $P^L_1(Q^2)$ sequence fit.}
\label{tab:syserror}
\end{table}

\section{{\boldmath $\eta$} transition form factor: a space- and time-like description}
\label{sec:PSTFF}

To extract the $\eta$ TFF low-energy parameters $b_{\eta}$, $c_{\eta}$, and $d_{\eta}$ (slope, curvature, and third derivative respectively) from the available data, we start with a $P^L_1(Q^2)$ sequence. However, according to Ref.~\cite{Lepage:1980fj}, the pseudoscalar TFFs behave as $1/Q^2$ for $Q^2\to\infty$, which means that, for any value of $L$, one will obtain in principle a good fit only up to a finite value of $Q^2$ but not for $Q^2\to\infty$. Therefore, it would be desirable to incorporate this asymptotic-limit information in the fits to $Q^2 F_{\eta \gamma^*\gamma}(Q^2)$ by considering also a $P^N_{N}(Q^2)$ sequence.

This method, which makes use of experimental data and theoretical framework for fitting them, cannot access the second Riemann sheet where the resonance poles are supposed to be located~\cite{Martin}.
One cannot extract resonance poles parameters with such methods, and that poses a word of caution on the interpretation of fits such as Eq.~(\ref{eq:tffpole}) to relate its pole parameters with effective masses. Our method does not contain a branch cut and all the analytical structure is built to reproduce only the first Riemann sheet. The effective pole we obtain should lie outside the range where data are.
The main advantage of the method of PAs is indeed to provide the $Q^2$ dependence of the TFF over the whole space- and time-like region up to the first resonance in an easy and systematic way, without the need of a model for the resonance poles appearing in the amplitude~\cite{Masjuan:2008fv,Masjuan:2012wy}. 
For how to extract resonance pole parameters using PA, see Refs.~\cite{Masjuan:2013jha,Masjuan:2014psa}.

Experimental data from the space-like region is obtained from CELLO, CLEO, and \babar \  Collaborations
\cite{Behrend:1990sr,Gronberg:1997fj,BABAR:2011ad}, together with the time-like experimental data from NA60 and A2 Collaborations \cite{Arnaldi:2009aa,Berghauser:2011zz,Aguar-Bartolome:2013vpw}. We also include the value $\Gamma_{\eta\to\gamma\gamma}=0.516(18)$~keV \cite{Agashe:2014kda} (which is basically dominated by the recent KLOE-2 measurement \cite{Babusci:2012ik}) in our fits.

\subsection{A remark on experimental systematic errors}

When comparing time-like data results from different collaborations it is common to report, together with the experimental data, the result of a fit with a single-pole function Eq.~(\ref{eq:tffpole}). Although such data contain only statistical errors, systematic errors are incorporated in the result of the fit. When using these data in our fits one must incorporate the systematic error information into the fitted data\footnote{We thank Marc Unverzagt for discussions along these lines.}. 

The A2 Collaboration reported in 2011 on the Dalitz decay $\eta \rightarrow e^+e^- \gamma$  \cite{Berghauser:2011zz}. Their fit yielded $\Lambda^{-2} = (1.92\pm 0.35_{stat} \pm 0.13_{syst}) $ GeV$^{-2}$. Combining both statistical and systematic error one obtains $\Lambda^{-2} = (1.92\pm 0.39_{comb}) $ GeV$^{-2}$. In order to obtain the combined error from a direct fit to the published data one can include a new source of error defined in the following way: $\Delta_{final}=\sqrt{ \Delta_{stat}^2 + (\epsilon |F(Q_i^2)|^2)^2}$ for each $Q^2_i$ datum, with $\epsilon$ a percentage. For the A2 2011 data we find that $\epsilon = 6.8\%$ will allow us to reproduce, with Eq.~(\ref{eq:tffpole}), the combined result $\Lambda^{-2} = (1.92\pm 0.39_{comb}) $ GeV$^{-2}$.

At the same time, an analysis of the $\eta \rightarrow \mu^+ \mu^- \gamma$ Dalitz decay by the NA60 Collaboration allowed a determination of $\Lambda^{-2}$ with significantly better statistical accuracy. In 2009, they reported the value $\Lambda^{-2} = (1.95\pm 0.17_{stat} \pm 0.05_{syst}) $ GeV$^{-2}$ \cite{Arnaldi:2009aa}\footnote{Recently, NA60 presented an improved preliminary result, $\Lambda^{-2} = (1.951\pm 0.059_{stat} \pm 0.042_{syst}) $ GeV$^{-2}$ \cite{Uras:2012qk} but the corresponding data are not yet published.},  which implies a factor $\epsilon =1.9\%$ for our $\Delta_{final}$.

In 2013, the A2 Collaboration reported a new measurement of the same Dalitz decay $\eta \rightarrow e^+e^- \gamma$ with larger statistics with a fitted value $\Lambda^{-2} = (1.95\pm 0.15_{stat} \pm 0.10_{syst}) $ GeV$^{-2}$ \cite{Aguar-Bartolome:2013vpw}, which leads to $\epsilon = 4.8\%$.
 
Published space-like data contains both error sources separately. The exception is the CELLO Collaboration which does not report a systematic error for each bin of data. Only a $12\%$ for the two-photon $\eta$-decay channel is reported. Accounting for all the different systematic sources we could find in their publication, we ascribe a $12\%$ of systematic error for the hadronic $\eta$ decay which leads to a $6\%$ error for the global number of events (implying a $12\%$ of systematic error for each bin). We expect that the forthcoming space-like measurements at BES-III will provide the accurate description of such energy region and the role of the unknown systematic effects in the CELLO data would not be important.

\subsection{Results}

\begin{figure*}[htbp]
\begin{center}
\includegraphics{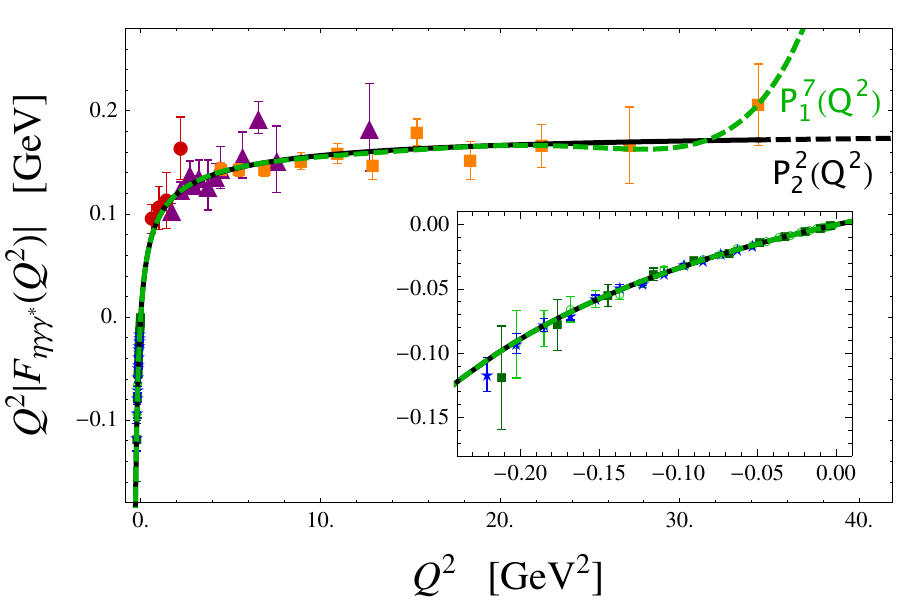}
\caption{$\eta$-TFF best fits. Green-dashed line shows our best $P^L_{1}(Q^2)$ fit and black line our best $P^N_N(Q^2)$ fit. Experimental data points in the space-like region are from CELLO (red circles) \cite{Behrend:1990sr}, CLEO (purple triangles) \cite{Gronberg:1997fj}, and \babar \ (orange squares) \cite{BABAR:2011ad} Collaborations. Experimental data points in the time-like region are from NA60 (blue stars)  \cite{Arnaldi:2009aa}, A2 2011 (dark-green squares) \cite{Berghauser:2011zz}, and A2 2013 (empty-green circles) \cite{Aguar-Bartolome:2013vpw}. The inner plot shows a zoom into the time-like region.}
\label{SLTLfit}
\end{center}
\end{figure*}

\begin{figure*}[htbp]
\begin{center}
\includegraphics[width=7.5cm]{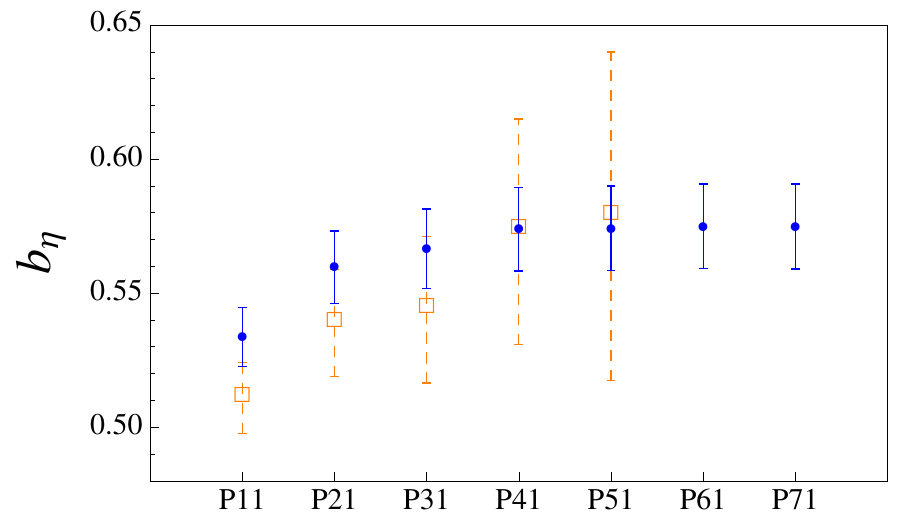}
\includegraphics[width=7.5cm]{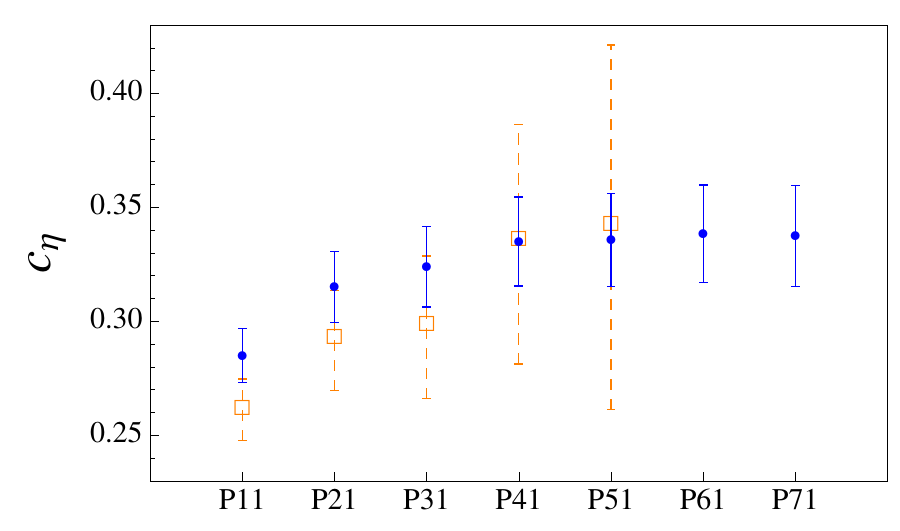}
\includegraphics[width=7.5cm]{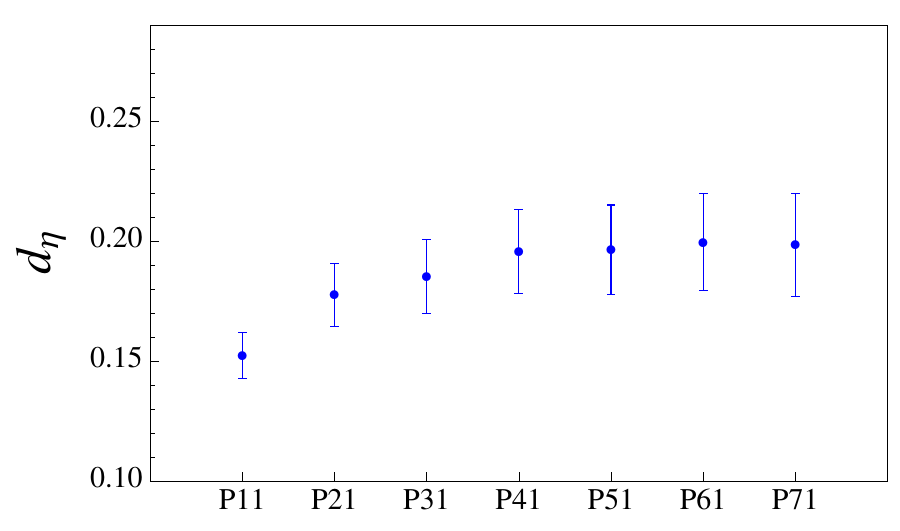}
\caption{Slope (top-left panel), curvature (top-right panel), and third derivative (bottom panel) predictions for the $\eta$ TFF using the $P^L_1(Q^2)$ up to $L=7$ (blue points). Previous results considering only space-like data from Ref.~\cite{Escribano:2013kba} are also shown (empty-orange squares) as a way to stress the role of the time-like data in our fits. Only statistical errors are shown.}
\label{fig:slope}
\end{center}
\end{figure*}

After defining the set of data we will use, we report on our results. We start fitting with a $P^L_1(Q^2)$ sequence. We reach $L=7$ and we show it in Fig.~\ref{SLTLfit} as a green-dashed line. The smaller plot in Fig.~\ref{SLTLfit} is a zoom into the time-like region. The obtained LEPs are collected in Table~\ref{tab:psresults} and shown in Fig.~\ref{fig:slope} together with our previous results (empty orange) when only space-like data were included in our fits~\cite{Escribano:2013kba}. The stability observed for the LEPs with the $P^L_1(Q^2)$ sequence is remarkable, and the impact of the inclusion of time-like data is clear since not only allows us to reach higher precision on each PA but also to enlarge our PA sequence by 2 elements. The stability of the result is also clearer and reached earlier, reduces our systematic error, and shows the ability of our method to extract, for the first time, the LEPs from a combined fit to all the available data.
The coefficients of the best fitted $P^L_1(Q^2)$ can be found in~\ref{AppTL}. 

\begin{table*}[ht]
\begin{center}
\begin{tabular}{|c||c|c|c|c|c|c|}
\hline
& \multicolumn{5}{c|}{$\eta$ TFF}\\
\hline
& $N$ & $b_{\eta}$ & $c_{\eta}$ & $d_{\eta}$ & $\chi^2$/dof \\
\hline
$P^N_1(Q^2)$ & $7$ & $0.575(16)$ & $0.338(22)$ & $0.198(21)$ & $0.6$ \\
$P^N_N(Q^2)$ & $2$ & $0.576(15)$ & $0.340(20)$ & $0.201(19)$ & $0.6$ \\
\hline
Final & $ $ & $0.576(11)$ & $0.339(15)$ & $0.200(14)$ & $ $ \\
\hline
\end{tabular}
\end{center}
\caption{Low-energy parameters for the $\eta$ TFF obtained from the PA fits to experimental data.
The first column indicates the type of sequence used for the fit and $N$ is its highest order.
The last row shows the weighted average result for each LEP. We also present the quality of the fits in terms of $\chi^2$/DOF (degrees of freedom). Errors are only statistical and symmetrical.}
\label{tab:psresults}
\end{table*}

To reproduce the asymptotic behavior of the TFF, we have also considered the $P^N_{N}(Q^2)$ sequence (second row in Table \ref{tab:psresults}). The results obtained are in very nice agreement with our previous determinations. The best fit is shown as black-solid line in Fig.~\ref{SLTLfit}. We reach $N=2$. Since these approximants contain the correct high-energy behavior built-in, they can be extrapolated up to infinity (black-dashed line in Fig.~\ref{SLTLfit}) and then predict the leading $1/Q^2$ coefficient:
\begin{equation}
\label{mixinglimits}
\begin{split}
\lim_{Q^2\to\infty }Q^2F_{\eta\gamma^*\gamma}(Q^2) &=0.177^{+0.020}_{-0.009}\ \textrm{GeV}\  .
\end{split}
\end{equation}

This prediction, although larger than in our previous work \cite{Escribano:2013kba}, still cannot be satisfactorily compared with the \babar \  time-like measurement  at $q^2=112$~GeV$^2$, $F_{\eta\gamma^*\gamma}(112 \textrm{ GeV}^2)=0.229(30)(8)$~GeV~\cite{Aubert:2006cy}. The impact of such discrepancy in the $\eta-\eta^\prime$ mixing is discussed in the next section. 

Our combined weighted average results from Table \ref{tab:psresults}, taking into account both types of PA sequences, give
\begin{equation}
\label{etaetapvalues}
\begin{split}
&\begin{cases}
 	&b_{\eta} = 0.576(11)_{\rm stat}(4)_{\rm sys}\\
 	&c_{\eta} = 0.339(15)_{\rm stat}(5)_{\rm sys}\\
	&d_{\eta} = 0.200(14)_{\rm stat}(18)_{\rm sys}
 \end{cases}
\end{split}
\end{equation}
where the second error is systematic (around $0.7$, $1.5$, and $9\%$ for $b_P$, $c_P$, and $d_P$, respectively, from Table~\ref{tab:syserror}).

Equation~(\ref{etaetapvalues}) can be compared with $b_{\eta} = 0.60(6)_{\rm stat}(3)_{\rm sys}$, $c_{\eta} = 0.37(10)_{\rm stat}(7)_{\rm sys}$ using space-like data exclusively~\cite{Escribano:2013kba}. As expected, not only statistical results have been improved but also systematics, both by an order of magnitude, yielding the most precise slope determination ever.

\begin{figure}[tbp]
\begin{center}
	\includegraphics[width=\columnwidth]{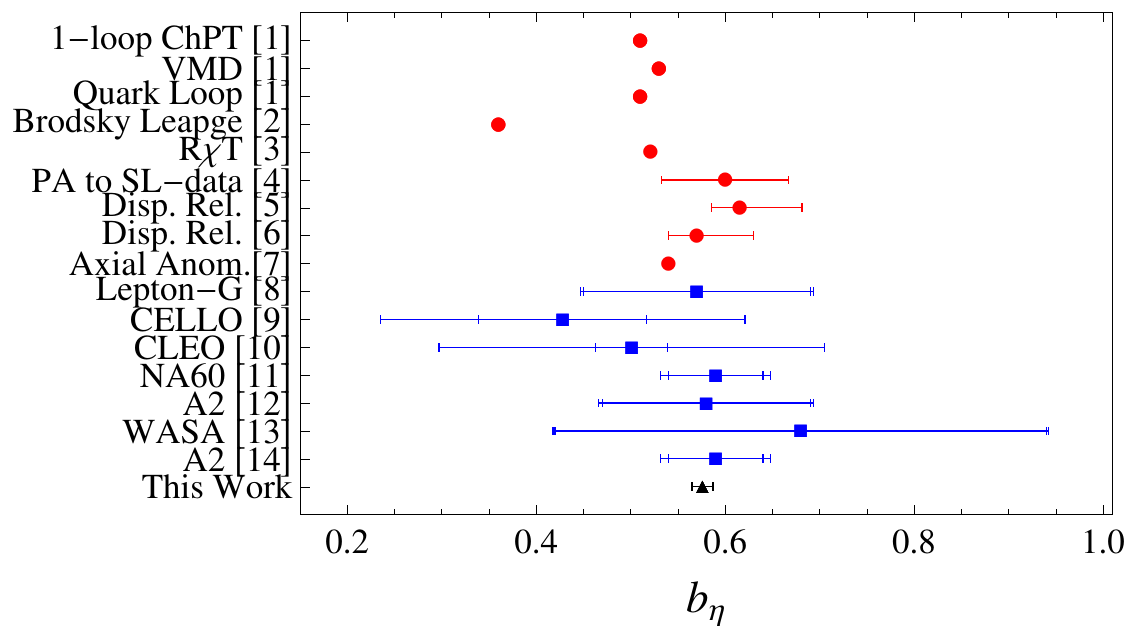}
\caption{Slope determinations for  $\eta$ TFF from different theoretical (red circles) and experimental (blue squares) references discussed in the text. Inner error is the statistical one and larger error is the combination of statistical and systematic errors.}
\label{fig:slopecomp}
\end{center}
\end{figure}

Our slope is compared with experimental determinations from~\cite{Dzhelyadin:1980kh,Behrend:1990sr,Gronberg:1997fj,Arnaldi:2009aa,Berghauser:2011zz,Hodana:2012rc,Aguar-Bartolome:2013vpw} together with theoretical extraction from~\cite{Ametller:1991jv,Brodsky:1981rp,Czyz:2012nq,Escribano:2013kba,Hanhart:2013vba,Kubis:2015sga,Klopot:2013laa} in Fig.~\ref{fig:slopecomp}.

One should notice that all the previous collaborations used a VMD model fit to extract the slope.
In order to be consistent when comparing with our results, a systematic error of about $40\%$ should be added to the experimental determinations based on space-like data \cite{Masjuan:2012wy,Escribano:2013kba}, and a systematic error of about $5\%$ should be added to the experimental determinations based on time-like data (see~\ref{AppConv} for further details).

When comparing different theoretical extractions of the slope of the $\eta$ TFF with our result in Fig.~\ref{fig:slopecomp}, we find a pretty good agreement with the exception of the results in Ref.~\cite{Czyz:2012nq} that reported $b_{\eta}=0.546(9)$ and $b_{\eta}=0.521(2)$ using Resonance Chiral Theory with one- or two-octet ans\"atze. The disagreement is between 2 and 5 standard deviations. Reference~\cite{Czyz:2012nq} uses Resonance Chiral Theory, which is based in large-$N_c$ arguments, to extract LEPs. Going from large-$N_c$ to $N_c=3$ imposes a systematic error~\cite{Masjuan:2007ay,Masjuan:2008fr,Queralt:2010sv,Masjuan:2012sk}. Since Ref.~\cite{Czyz:2012nq} considered two approximations for fitting the $\eta$ TFF (with one and two octets), one could consider the difference between them as a way to estimate such error~\cite{Masjuan:2012qn,Masjuan:2013jha,Escribano:2013kba}. In such a way, the $\eta$ TFF slope would read $b_{\eta}=0.53(1)$, at 2.5 standard deviation from our result.

Eventually, we want to comment on the effective single-pole mass determination $\Lambda_P$ from Eq.~(\ref{eq:tffpole}).
Using $b_P=m_P^2/\Lambda_{P}^2$ and the values in Eq.~(\ref{etaetapvalues}),
we obtain $\Lambda_{\eta}=0.722(7)$ GeV or $\Lambda_{\eta}^{-2}=1.919(39)$ GeV$^{-2}$.

The fits shown in Fig.~\ref{SLTLfit} use the experimental value of the two-photon decay width as an experimental datum to be fitted. Such fit could be repeated without including that decay. In such a way, we reach again a $P^7_1(Q^2)$ and a $P^2_2(Q^2)$ as our best PA with the advantage now that the value $F_{\eta \gamma \gamma}(0)$ is a prediction of our fits. We find $F_{\eta \gamma \gamma}(0)|_{fit}=0.250(38)$~GeV$^{-1}$ for the  $P^7_1(Q^2)$ and $F_{\eta \gamma \gamma}(0)|_{fit}=0.248(28)$~GeV$^{-1}$ for the $P^2_2(Q^2)$, which translates into $\Gamma_{\eta\gamma\gamma}|_{fit}=0.43(13)$~keV and $\Gamma_{\eta\gamma\gamma}|_{fit}=0.42(10)$~keV, respectively. Comparing with the experimental value $\Gamma_{\eta\gamma\gamma}|_{exp}=0.516(18)$~keV such predictions are at 0.66 and 0.94 standard deviation each. 

\subsection[The impact of $\eta\rightarrow\gamma\gamma$ measurements]{The impact of {\boldmath $\eta\rightarrow\gamma\gamma$} measurements}

Our results in Eq.~\eqref{etaetapvalues} are, by far, the most precise to date. Particularly, we believe that the precision achieved for $b_{\eta}$ will be hard to improve even if new data becomes available. Nevertheless, the values obtained mildly depend on $\Gamma_{\eta\gamma\gamma}$. For instance, if we would have used the value measured through the Primakoff mechanism  
omitted in the PDG average~\cite{Agashe:2014kda} (i.e, $\Gamma_{\eta\gamma\gamma}^{\mathrm{Primakoff}} = 0.476(62)$~keV~\cite{Rodrigues:2008zza}), we would find  $b_{\eta}=0.570(13)$ for the SL+TL extraction with a $20\%$ larger error, still in nice agreement with our aforementioned results. Notice that this result is pretty similar to the one obtained by our fits when the decay into two photons is not used in the data set. This fact does not draw a puzzle, everything seems to agree within uncertainties, but may suggest to look again for a Primakoff measurement.

\subsection{The role played by high-energy space-like data}

Low-energy parameters are defined at zero momentum transfer. When extracting them from our fits, one would expect the low-energy data to dominate. We noticed, however, that in order to reach large PA sequences (leading to more precise extractions), the high-energy data is also important as can be seen in Fig.~\ref{SLTLfit}. From $5$~GeV$^2$ to $35$~GeV$^2$ data are basically dominated by the \babar \  measurement~\cite{BABAR:2011ad}  and that has a clear implication on the extraction of the asymptotic $\lim_{Q^2\to\infty }Q^2F_{\eta\gamma^*\gamma}(Q^2)$ value as can be seen in Table~\ref{tab:coll}, where the role of data from each collaboration is reported. Indeed, a fit  exclusively to \babar \  data yields similar results for both slope and asymptotic value than when considering the full set of space-like data. However, a fit to the data from the CELLO~\cite{Behrend:1990sr} Collaboration which range only up to $2.23$ GeV$^2$ yields much larger asymptotic value (although statistically compatible). Considering only data from the CLEO~\cite{Gronberg:1997fj} Collaboration, which ranges up to $12.74$ GeV$^2$, reduces the asymptotic value by about $20\%$ compared to CELLO. The role of \babar \  data is then twofold, allowing first to reach $N=2$ in the $P^N_N(Q^2)$ sequence and determining basically the asymptotic value.

In view of the puzzle of the $\pi^0$ TFF between \babar \ ~\cite{Aubert:2009mc} and Belle~\cite{Uehara:2012ag} results, a second experimental measurement covering the high-energy region would be very welcome here. We find the Belle Collaboration suited for such purpose and we would like to encourage them to go ahead with such measurements.

On the other side, time-like data can also be used to predict the asymptotic value, even though the range of data is much shorter and much closer to $Q^2=0$. From the three sets of time-like data used in our fits, A2-11~\cite{Berghauser:2011zz}  and A2-13~\cite{Aguar-Bartolome:2013vpw}  are based on the $\eta\to e^+e^-\gamma$ and covers larger range of phase space. The NA60~\cite{Arnaldi:2009aa} Collaboration, based on  the $\eta \to \mu^+ \mu^- \gamma$, covers a shorter range but in the higher-energy region. The asymptotic values extracted from the difference time-like sets of data agree rather well but disagree with the results obtained from the space-like data (although overlapping within errors). Whatever the combination of different data sets selected, \babar \  data always decides on the asymptotic value. In passing, we notice that any of the configurations considered so far agrees with the results of the $\eta$ TFF measurement from \babar \ ~\cite{Aubert:2006cy}.

\begin{table*}
\begin{center}
\begin{tabular}{|c|c||c|c||c|c|c|}
\hline
& data range & \multicolumn{2}{c||}{$P^L_1(Q^2)$} & \multicolumn{3}{c|}{$P^N_N(Q^2)$} \\
\cline{3-7}
& (GeV$^2$) & $L$ & $b_{\eta}$ & $N$ & $b_{\eta}$ &  $\lim\limits_{Q^2\to\infty }Q^2F_{\eta\gamma^*\gamma}(Q^2)$ \\
\hline
CELLO~\cite{Behrend:1990sr} & 0.62--2.23  &$2$ & $0.48(20)$ & $1$ & $0.427(66)$ & $0.193(30)$ \\
CLEO~\cite{Gronberg:1997fj} & 1.73--12.74&$3$ & $0.73(12)$ & $1$ & $0.522(19)$ & $0.157(5)$ \\
\babar~\cite{BABAR:2011ad} & 4.47--34.38&$4$ & $0.53(9)$ & $1$ & $0.509(14)$ & $0.162(3)$ \\
CELLO+CLEO~\cite{Behrend:1990sr,Gronberg:1997fj} &0.62--12.74 & $3$ & $0.65(9)$ & $2$ & $0.704(87)$ & $0.25(10)$ \\
SL &0.62--34.38 & $5$ & $0.58(6)$ & $2$ & $0.66(10)$ & $0.161(24)$ \\
\hline
A2-11+A2-13~\cite{Berghauser:2011zz,Aguar-Bartolome:2013vpw}  & -0.212 -- -0.002& $2$ & $0.475(76)$ & $1$ & $0.551(40)$ & $0.149(11)$ \\
NA60~\cite{Arnaldi:2009aa} & -0.221 -- -0.053 & $3$ & $0.640(77)$ & $1$ & $0.582(19)$ & $0.141(5)$ \\
TL  & -0.221 -- -0.002& $3$ & $0.565(87)$ & $1$ & $0.576(17)$ & $0.143(5)$ \\
\hline
CELLO~\cite{Behrend:1990sr} + TL & -0.221 -- 2.23 &$5$ & $0.531(39)$ & $2$ & $0.533(30)$ & $0.203(58)$ \\
CELLO+CLEO~\cite{Behrend:1990sr,Gronberg:1997fj} +TL &  -0.221 -- 12.74 & $6$ & $0.567(22) $ & $1$ & $0.550(13)$ & $0.152(3)$ \\
\hline
A2-11+A2-13~\cite{Berghauser:2011zz,Aguar-Bartolome:2013vpw} +SL &   -0.212 -- 34.38 & $7$ & $0.561(35)$ & $2$ & $0.569(28)$ & $0.178(16)$ \\
\hline
\textbf{TL +SL} & \textbf{ -0.221 -- 34.38}  & $\mathbf{7}$ & $\mathbf{0.575(16)}$ & $\mathbf{2}$ & $\mathbf{0.576(15)}$ & $\mathbf{0.177(15)}$ \\
\hline
\end{tabular}
\end{center}
\caption{Role of the different sets of experimental data in determining slope and asymptotic values of the $\eta$ TFF. \textit{SL} refers the the space-like data set, i.e., data from CELLO+CLEO+\babar \ ~\cite{Behrend:1990sr,Gronberg:1997fj,BABAR:2011ad} Collaborations, and \textit{TL} refers to the time-like data set, i.e., data from NA60+A2-11+A2-13~\cite{Arnaldi:2009aa,Berghauser:2011zz,Aguar-Bartolome:2013vpw} Collaborations. Bold numbers are our final result. No systematic errors included.}
\label{tab:coll}
\end{table*}

\section{{\boldmath $\eta$} transition form factor: applications}
\label{sec:mixing}

As sated in the introduction, TFF are not also interesting by themselves but also for the range of scenarios where they play a crucial role. In this section we consider few of such applications.

\subsection[Reanalysis of the $\eta$-$\eta^\prime$ mixing parameters]{Reanalysis of the {\boldmath $\eta$-$\eta^\prime$} mixing parameters}

In this subsection we briefly summarize the main elements to extract the mixing parameters exclusively from our fits to the form factor data.

As was done in Ref.~\cite{Escribano:2013kba}, we analyze $\eta$-$\eta^\prime$ mixing using the quark-flavor basis.
In this basis, the $\eta$ and $\eta^\prime$ decay constants are parametrized as
\begin{equation}
   \left(
      \begin{array}{cc}
          F^{q}_{\eta} & F^{s}_{\eta}\\
          F^{q}_{\eta^\prime} & F^{s}_{\eta^\prime}\\
      \end{array}
   \right)
=
   \left(
      \begin{array}{cc}
          F_{q}\cos\phi_{q} & -F_{s}\sin\phi_{s}\\
          F_{q}\sin\phi_{q} & F_{s}\cos\phi_{s}\\
      \end{array}
   \right)\ ,
\end{equation}
where $F_{q,s}$ are the light-quark and strange pseudoscalar decay constants, respectively,
and $\phi_{q,s}$  the related mixing angles.
Several phenomenological analyses find $\phi_q\simeq\phi_s$, 
which is also supported by large-$N_c$ ChPT calculations
where the difference between these two angles is seen to be proportional to an OZI-rule violating parameter and hence small \cite{Feldmann:1998vh,Escribano:2005qq}.

Within this approximation, the asymptotic limits of the TFFs take the form 
\begin{equation}
\label{eq:FLAVEtaAsymp}
\begin{split}
\lim_{Q^2\to\infty }Q^2F_{\eta\gamma^*\gamma}(Q^2) & =2(\hat c_q F^q_{\eta}+\hat c_s F^s_{\eta}) \\
& =2(\hat c_q F_q\cos\phi-\hat c_s F_s\sin\phi)\ ,\\[1ex]
\lim_{Q^2\to\infty }Q^2F_{\eta^\prime\gamma^*\gamma}(Q^2) & =2(\hat c_q F^q_{\eta^\prime}+\hat c_s F^s_{\eta^\prime})\\
& =2(\hat c_q F_q\sin\phi+\hat c_s F_s\cos\phi)\ ,
\end{split}
\end{equation}
and their normalization at zero (from the chiral anomaly and Eq.~\eqref{eq:tffdecay})
\begin{equation}
\label{eq:FLAVEtaDecay}
\begin{split}
F_{\eta\gamma\gamma}(0) & =
\frac{1}{4\pi^2}\left(\frac{\hat c_q F^s_{\eta^\prime}-\hat c_s F^q_{\eta^\prime}}{F^s_{\eta^\prime}F^q_{\eta}-F^q_{\eta^\prime}F^s_{\eta}}\right) \\
 & =\frac{1}{4\pi^2}\left(\frac{\hat c_q}{F_q}\cos\phi-\frac{\hat c_s}{F_s}\sin\phi\right)\ ,\\[1ex]
F_{\eta^\prime\gamma\gamma}(0) & =
\frac{1}{4\pi^2}\left(\frac{\hat c_q F^s_{\eta}-\hat c_s F^q_{\eta}}{F^q_{\eta}F^s_{\eta^\prime}-F^s_{\eta}F^q_{\eta^\prime}}\right) \\
 & =\frac{1}{4\pi^2}\left(\frac{\hat c_q}{F_q}\sin\phi+\frac{\hat c_s}{F_s}\cos\phi\right)\ ,
\end{split}
\end{equation}
with $\hat c_q=5/3$ and $\hat c_s=\sqrt{2}/3$.

Experimental information provides $|F_{\eta\gamma\gamma}(0)|_{\rm exp}=0.274$ $(5)$ GeV$^{-1}$ and $|F_{\eta^\prime\gamma\gamma}(0)|_{\rm exp}=0.344(6)$ GeV$^{-1}$
and for the asymptotic value of the $\eta$ TFF we take the value shown in Eq.~(\ref{mixinglimits}) with symmetrical errors, $\lim_{Q^2\to\infty }Q^2F_{\eta\gamma^*\gamma}(Q^2)=0.177(15)$ GeV.
With these values, the mixing parameters are predicted to be 
\begin{equation}
\label{eq:mixresults}
F_q/F_\pi=1.07(1)\ ,\quad
F_s/F_\pi=1.39(14)\ , \quad
\phi=39.3(1.2)^{\circ}\ ,
\end{equation}
with $F_\pi=92.21(14)$ MeV \cite{Agashe:2014kda}. The uncertainties are dominated by the error from the asymptotic value prediction. 

One can translate the mixing parameters obtained in the flavor \ bases \  into \  the octet-singlet one by the following recipe~\cite{Feldmann:1998sh}:
\begin{equation}
\label{dictionary}
\centering
\begin{split}
F_8^2=\frac{F_q^2+2 F_s^2}{3}\, , &\quad F_0^2 = \frac{2 F_q^2 + F_s^2}{3}\, ,\\
\theta_8=\phi - \arctan(\frac{\sqrt{2} F_s}{F_q})\, ,&\quad  \theta_0=\phi - \arctan(\frac{\sqrt{2} F_q}{F_s})\, .
\end{split}
\end{equation}
where
\begin{equation}
   \left(
      \begin{array}{cc}
          F^{8}_{\eta} & F^{0}_{\eta}\\
          F^{8}_{\eta^\prime} & F^{0}_{\eta^\prime}\\
      \end{array}
   \right)
=
   \left(
      \begin{array}{cc}
          F_{8}\cos\theta_{8} & -F_{0}\sin\theta_{0}\\
          F_{8}\sin\theta_{8} & F_{0}\cos\theta_{0}\\
      \end{array}
   \right)\ ,
\end{equation}
represents the admixture of the $\eta$ and $\eta'$ decay constants in terms of the octet and singlet one.

These relations, Eqs.~\eqref{dictionary}, are very useful since, as observed in Ref.~\cite{Leutwyler:1997yr} and recently discussed in~\cite{Agaev:2014wna}, the singlet decay constant $F_0$ is renormalisation-scale dependent:
\begin{equation}
\label{eq:F0running}
\begin{array}{rl}
\mu\frac{d F_0}{d\mu} & = - N_F \left(\frac{\alpha_s(\mu)}{\pi}\right)^2 F_0\\[1ex]
\longrightarrow \quad F_0(\mu) & = F_0(\mu_0)\left(1 + \frac{2N_F}{\beta_0}\left(  \frac{\alpha_s(\mu)}{\pi} -\frac{ \alpha_s(\mu_0)}{\pi}  \right)\right)\\
 & = F_0(\mu_0)\left(1 + \delta \right)\, ,
\end{array}
\end{equation}
\noindent
with $\beta_0 = \frac{11N_c}{3}-\frac{2}{3}N_F$, $N_c$ the number of colors, $N_F$ the number of active flavors at each scale, and $\mu$ the renormalization-scale, with $\mu_0=1$ GeV a reference point close to the $\eta'$ mass. 

To include this effect in our results it is convenient to work it out in the singlet-octet basis for later on translate it into the 
flavor one using~\eqref{dictionary}. As such, 
the asymptotic behaviour Eqs.~\eqref{eq:FLAVEtaAsymp} shift to
\begin{equation}
\label{asympdelta}
\begin{split}
& \lim_{Q^2\to\infty }Q^2F_{\eta\gamma^*\gamma}(Q^2)  =   \\ & \qquad 2(\hat c_q(1+4\delta/5) F_q\cos\phi-   \hat c_s(1 + 2\delta) F_s\sin\phi),  \\ 
& \lim_{Q^2\to\infty }Q^2F_{\eta^\prime\gamma^*\gamma}(Q^2)   =  \\ & \qquad 2(\hat c_q (1+4\delta/5)F_q\sin\phi+  \hat c_s (1 + 2\delta)F_s\cos\phi)\, .
\end{split}
\end{equation}

Assuming asymptotic freedom for $\alpha_s(\mu)$, the phenomenological input $\alpha_s(M_z)=0.1185$
\cite{Agashe:2014kda}, and the renormalisation group equation for $\alpha_s(\mu)$, we determine $\alpha_s(\mu_0=1\textrm{GeV})=0.48$, including up to four loop
corrections and threshold effects for its running\footnote{Particular details of the $\alpha_s$ running are irrelevant at the precision we are working.}. With such values and Eq.~\eqref{eq:F0running} we determine
$\delta = -0.17$. Using \eqref{dictionary} to go back to the flavor basis we obtain as our final mixing parameters, representing one of the main results of this work:
\begin{equation}
\label{eq:mixingetap}
\begin{array}{l}
{\bf inputs:}\, F_{\eta \gamma\gamma}(0),  F_{\eta' \gamma\gamma}(0), \mathrm{asymp }\,  \eta\, \\[1ex]
\Rightarrow \ F_q/F_\pi=1.07(2) , \  F_s/F_\pi=1.29(16) ,\ \phi=38.3(1.6)^{\circ},\\[2ex]
{\bf inputs:}\, F_{\eta \gamma\gamma}(0),  F_{\eta' \gamma\gamma}(0), \mathrm{asymp }\,  \eta'\,  \\[1ex]
\Rightarrow \   F_q/F_\pi=1.06(1) , \  F_s/F_\pi=1.63(8) ,\ \phi=41.1(0.8)^{\circ} ,
\end{array}
\end{equation}
when taking the $\eta(\eta')$ asymptotic behavior respectively as part of the subset of equations to be solved (\ref{eq:FLAVEtaDecay},\ref{asympdelta}).
We stress that corrections from $\delta$ are bigger for the $\eta'$ case, as the singlet admixture is more important there. 
Comparing to our old results (sec. III in~\cite{Escribano:2013kba}), we find a better agreement among both solutions in \eqref{eq:mixingetap}. 
As it was explained in Ref.~\cite{Escribano:2013kba}, the mixing Eqs. are not independent, there is a relation among them:
\begin{equation}
\label{eq:degenRUN}
\begin{split}
\lim_{Q^2\to\infty }Q^2\left( F_{\eta\gamma^*\gamma}(Q^2)F_{\eta\gamma\gamma}(0) +  F_{\eta'\gamma^*\gamma}(Q^2) F_{\eta'\gamma\gamma}(0) \right) = \\ \left(1+\frac{8}{9}\delta \right)\frac{3}{2\pi^2} ,
\end{split}
\end{equation}
where in the last step we have used the exact expressions (\ref{eq:FLAVEtaDecay},\ref{asympdelta}). Numerically, using our $\delta=-0.17$, one would obtain $0.85\frac{3}{2\pi^2}$ for the r.h.s. of \eqref{eq:degenRUN}. Our numerical predictions for the asymptotic form factors together with their experimental normalization yield $0.89(3) \frac{3}{2\pi^{2}}$ for its l.h.s., in nice agreement, but this would not be the case without the $\delta$ correction.

\begin{figure}[tbp]
\begin{center}
	\includegraphics[width=\columnwidth]{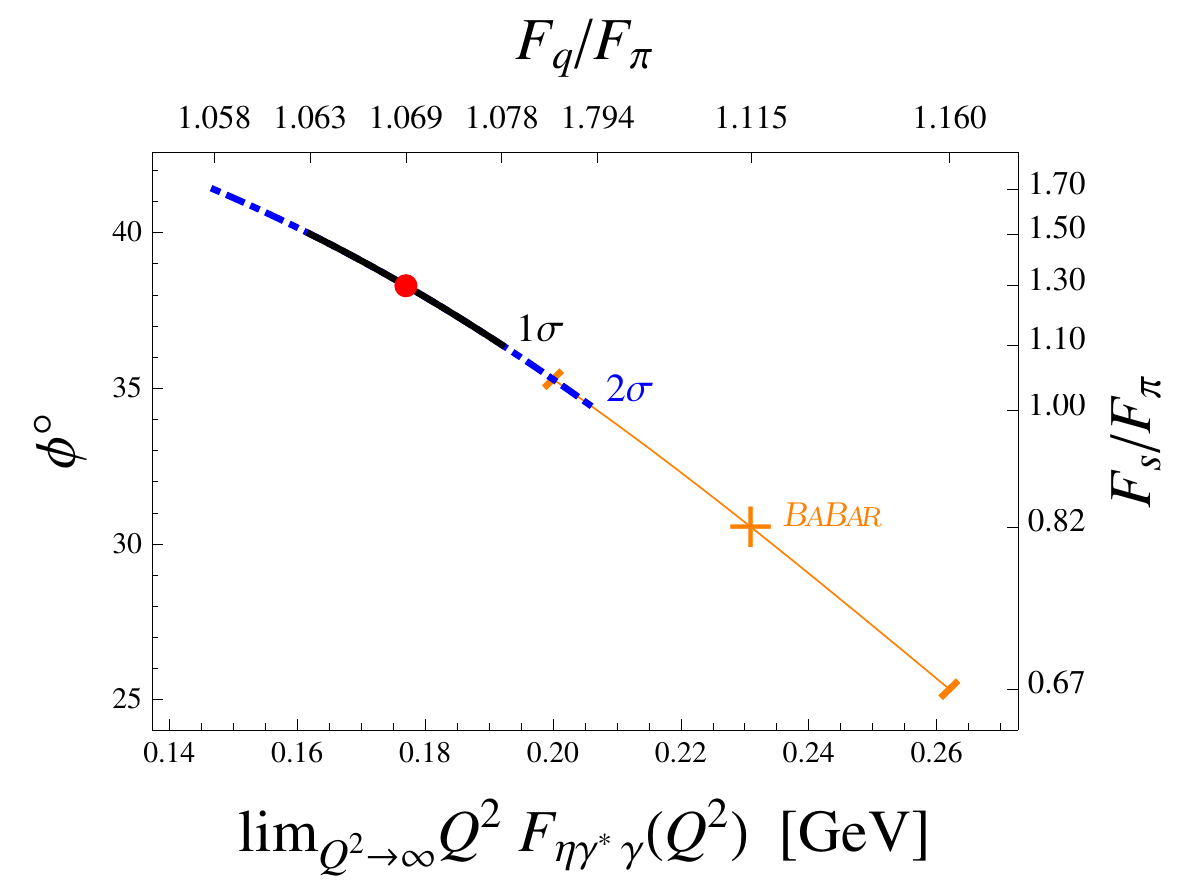}
\caption{Mixing parameters as a function of the $\eta$ TFF asymptotic value when the errors coming from the normalization of the TFFs is set to zero. For comparison, we show the mixing parameters extracted from the measurement of the time-like $\eta$ TFF at $q^2=112$~GeV$^2$ by \babar \  Collaboration~\cite{Aubert:2006cy}. This figure exemplifies the puzzle between the standard mixing parameters and the \babar \  measurement.}
\label{fig:plotmixing}
\end{center}
\end{figure}

This result contrast with \babar \  determinations, which, taking the running from $\mu_0$ up to their scale $Q^2=112$~GeV$^2$ instead of at $\infty$ (resulting in $\delta_{\mathrm{\footnotesize{\babar}}}=-0.09$), yields 
\begin{equation}
\label{babarmixing}
\begin{array}{c}
F_q/F_\pi=1.10(3)\ , \ \  F_s/F_\pi=0.91(21)\ ,\ \ \phi=33(4)^{\circ}, \\[1ex]
F_q/F_\pi=1.08(2)\ , \ \  F_s/F_\pi=1.17(23)\ ,\ \ \phi=37(3)^{\circ},
\end{array}
\end{equation}
using again the two subsets of equations in (\ref{asympdelta}). Eq.~\eqref{eq:degenRUN}, l.h.s., would read $0.98(7)\frac{3}{2\pi^2}$ where we are neglecting any $1/Q^2$ dependence in it. In other words, assuming that $Q^2=112$GeV$^2$ plays the role of $\infty$. The r.h.s of \eqref{eq:degenRUN} using $\delta_{\mathrm{\footnotesize{\babar}}}$ results in $0.92\frac{3}{2\pi^2}$, then compatible with its l.h.s. This comparison is somewhat false since we do not assume $\delta_{\mathrm{\footnotesize{\babar}}}$ to be at $\infty$, otherwise we would have used $\delta=-0.17$, and then we would have found a contradiction between l.h.s. and r.h.s. of \eqref{eq:degenRUN}. This discrepancy is what we call the \babar \  puzzle; it is depicted in Fig.~\ref{fig:plotmixing}. This figure includes both our mixing results and the \babar \  determination at $Q^2=112$GeV$^2$
with the corresponding mixing parameters obtained using Eqs.~(\ref{asympdelta},\ref{eq:FLAVEtaDecay}). 

We remark that our results are tied to $\alpha_s^2$ corrections in~\eqref{eq:F0running} and non-negligible systematic effects for the $\eta'$ asymptotic behavior from our fits.
Since this assertion relies on the high energy TFF-behavior, where only \babar \  data are available, a second measurement by Belle Collaboration would be a very useful crosscheck.

\begin{figure*}[htb]
\begin{center}
	\includegraphics[width=0.39\textwidth]{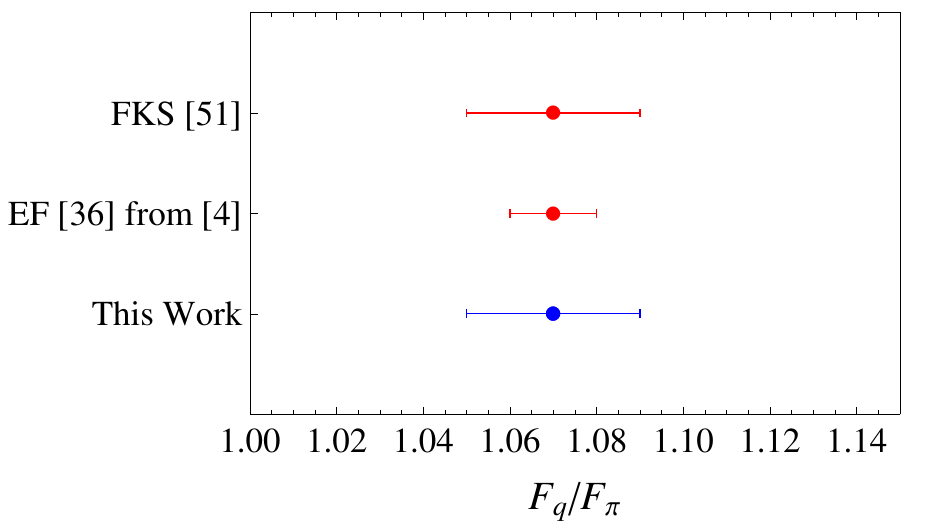}
	\includegraphics[width=0.29\textwidth]{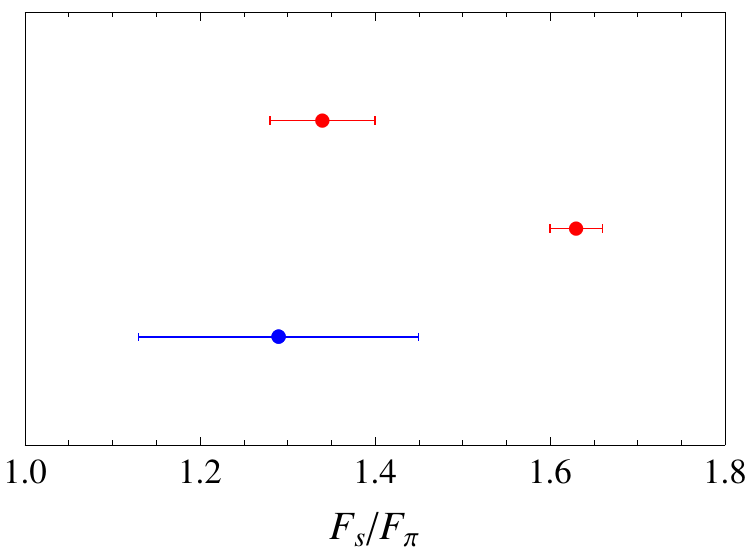}
	\includegraphics[width=0.29\textwidth]{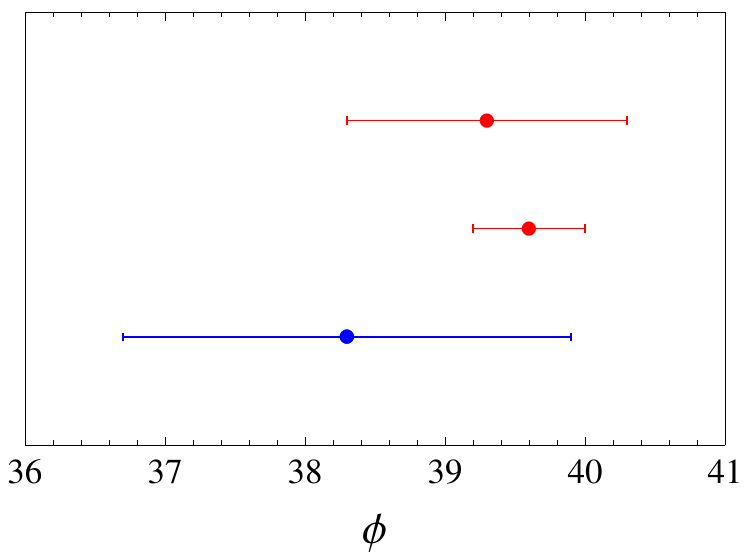}
\caption{Mixing parameters of the $\eta-\eta^\prime$ system in the flavor basis from different references.}
\label{fig:mixingflavor}
\end{center}
\end{figure*}

The mixing parameters obtained with our fits are precise enough to be competitive with the standard approaches with the advantage of using much less input information. Fig.~\ref{fig:mixingflavor} compares our results from Eq.~(\ref{eq:mixingetap}) (blue squares) with well-established phenomenological determinations, the one from Feldmann, Kroll, and Stech (FKS) from Ref.~\cite{Feldmann:1998vh,Feldmann:1998sh}, and the one from Escribano and Frere (EF) from Ref.~\cite{Escribano:2005qq} (updated in Ref.~\cite{Escribano:2013kba}). The agreement among the three approaches for both $F_q$ and $\phi$ is impressive. Less agreement is found for $F_s$. This parameter is more sensible to meson decays where the strange quark plays an important role, such as the $\phi$-meson decays. In fact, such decays where included in the EF approach but not in the FKS or in the present work.

\begin{figure*}[htb]
\begin{center}
	\includegraphics[width=0.4\textwidth]{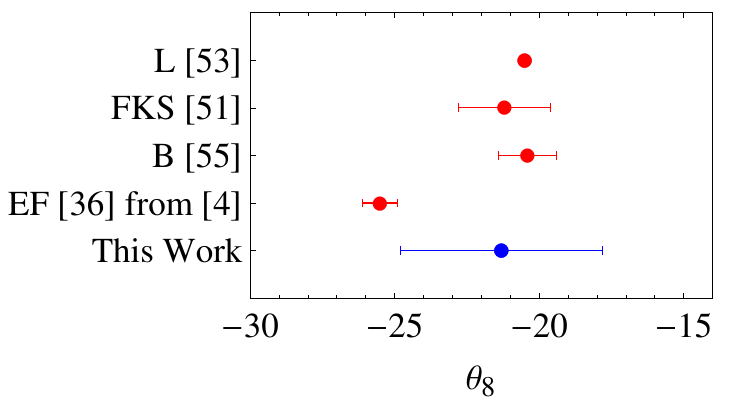}
	\hspace{-2.5cm}\includegraphics[width=0.4\textwidth]{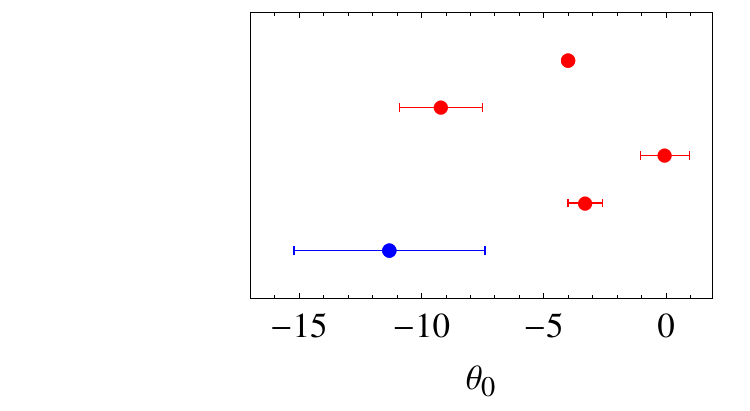}\\
	\includegraphics[width=0.4\textwidth]{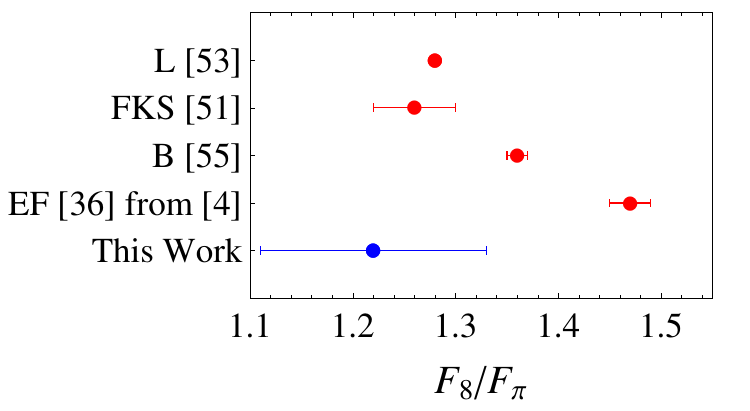}
	\hspace{-2.5cm}\includegraphics[width=0.4\textwidth]{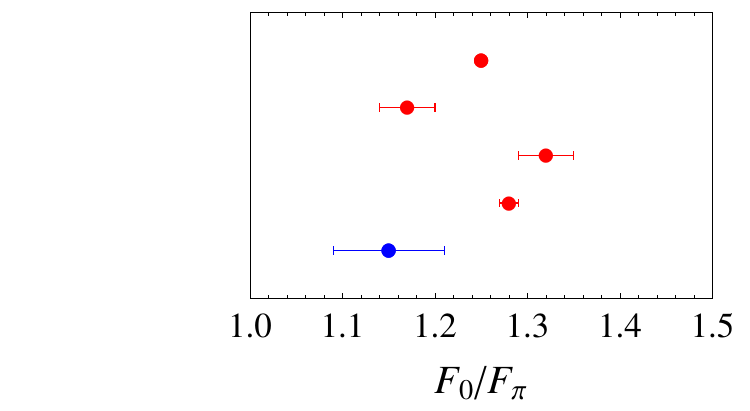}
\caption{Mixing parameters of the $\eta-\eta^\prime$ system in the octet-singlet basis from different references.}
\label{fig:mixing80}
\end{center}
\end{figure*}

Fig.~\ref{fig:mixing80} compares different singlet octet determinations. In this basis, our results turn out to be 
\begin{equation}
\label{08results}
\begin{array}{ll}
F_8/F_{\pi}=1.22(11)\, , &\qquad F_0/F_{\pi}=1.15(6)\, ,\\[1ex]
\theta_8 =-21.3(3.5)^\circ\, , &\qquad \theta_0=-11.3(3.9)^\circ\, ,
\end{array}
\end{equation}
and they can be compared with the FKS and EF as before together with the results by Leutwyler (L) from Ref.~\cite{Leutwyler:1997yr} (no errors were given), and the results from Benayoun, DelBuono, and O'Connell (BDO)~\cite{Benayoun:1999au}. Again, the agreement between our results and FKS and EF are remarkable, and also in agreement with the results of Leutwyler. Our results slightly disagree for $\theta_0$ and $F_0$ with BDO. The reason is because in the BDO the OZI violating piece is not set to zero. Since such piece mixes with the singlet component of the mixing, their $\theta_0$ and $F_0$ are slightly shifted.

\subsection{A comment on the \babar \  high-energy time-like measurement}

Our mixing parameters~\eqref{eq:mixingetap} disagree with the ones obtained using \babar \  time-like result as shown in Fig.~\ref{fig:plotmixing} 
 which, including the results in Figs.~\ref{fig:mixingflavor} and \ref{fig:mixing80} suggests a puzzle between 
the \babar \  measurement and the standard phenomenology~\cite{Escribano:2013kba}. Assuming that the measurement is correct, the difference could be explained if the assumption of duality between time- and space-like regions at high energy would not be yet valid at $112$~GeV$^2$ or that the asymptotic limit of Brodsky and Lepage is not yet reached at such energies and a systematic error is done in assuming duality~\cite{Aubert:2006cy}.

In Ref.~\cite{Aubert:2006cy}, \babar \  collaboration studied the process $e^+e^- \ \to \ \gamma^* \ \to \ \eta^{(')} \gamma$ at the center-of-mass energy $\sqrt{s}=10.58$ GeV. They measured its cross section and using its relation with the TFF, obtained the absolute value of the time-like TFF at $Q^2=-s=-112$GeV$^2$, 
$|Q^2 F_{\eta\gamma^*\gamma}(Q^2)|=(0.229\pm0.031)$ GeV  and  
$|Q^2 F_{\eta'\gamma^*\gamma}(Q^2)|=(0.251\pm0.021)$ GeV, where statical and systematic uncertainties are added in quadrature. 

A kinematic factor $K_P^3$ with $K_P=1-\frac{M_P^2}{s}$ (see~\cite{Rosner:2009bp}) was missing in \babar \  expressions. This correction leaves \babar \  published results almost untouched. This small shift together with the duality argument~\cite{Aubert:2006cy}, results in a prediction of the TFF at $Q^2=112$~GeV$^2$:
\begin{equation}
\label{BABARSL}
\begin{split}
|Q^2 F_{\eta\gamma^*\gamma}(Q^2)|_{Q^2=+112\mathrm{GeV}^2}=(0.231\pm0.031)~\mathrm{GeV}\, ,\\
|Q^2 F_{\eta'\gamma^*\gamma}(Q^2)|_{Q^2=+112\mathrm{GeV}^2}=(0.254\pm0.021)~\mathrm{GeV}\, .
\end{split}
\end{equation}

One is tempted to include these time-like measurements transformed into space-like predictions~\eqref{BABARSL} into our fits, after assuming that at this high momentum transfer, the duality between time- and space-like holds and no extra error should be included. For the $\eta$ TFF fits, its inclusion will mainly modify the asymptotic prediction growing up its value up to $\lim_{Q^2\to\infty }Q^2F_{\eta\gamma^*\gamma}(Q^2)=0.247$~GeV, higher than the \babar \  result, with a good reduced $\chi^2<1$. This, by itself, already indicates that at $Q^2=112$~GeV$^2$ the asymptotic regime is not yet reached. Curiously enough, the value of the fit function at $Q^2=-112$~GeV$^2$ is $|Q^2 F_{\eta\gamma^*\gamma}(Q^2)|_{Q^2=+112\mathrm{GeV}^2}=0.219$~GeV, below \eqref{BABARSL}. Even worse is the prediction of our fit function for the time-like counterpart, i.e., $|Q^2 F_{\eta\gamma^*\gamma}(Q^2)$ $|_{Q^2=-112\mathrm{GeV}^2} =0.307$~GeV. This exercise shows that the assumption of asymptotic regime at $112$GeV$^2$ has an error of about $15\%$ in our fits, a theoretical error that should be added to \babar \  results when used in the space-like region. A recent analysis of the pseudoscalar TFF based on perturbative corrections \cite{Agaev:2014wna} concludes that the difference between the time- and space-like form factors at $|Q^2|=112$~GeV$^2$ can be of the order of $5\%$ to $13\%$ for different pseudoscalar distribution amplitudes, and can be enhanced by Sudakov-type corrections (see~\cite{Bakulev:2000uh} for details). The Regge model defined in Ref.~\cite{Escribano:2013kba} also suggests a departure from duality of about $15\%$ to $20\%$ at $|Q^2|=112$~GeV$^2$. It is, however, difficult to calculate that error and hence difficult to ascribe it to the \babar \  determination.

Interestingly enough, to check the eventual departure arising from duality violations, one could artificially enhance  \babar \  error just to cross-check its order of magnitude. Increasing the error in Eq.~\eqref{BABARSL} from $0.031$~GeV$^2$ to $0.051$~GeV$^2$ (adding in quadrature a $1.3\sigma$) and refitting again, we obtain, whit somewhat better $\chi^2$, that the asymptotic predicted value would then be $\lim_{Q^2\to\infty }Q^2F_{\eta\gamma^*\gamma}(Q^2)=0.193$~GeV, the fit value at $Q^2=-112$~GeV$^2$ would read $0.187$~GeV but also our time-like prediction at $Q^2=112$~GeV$^2$ would read $0.199$~GeV, essentially satisfying the initial assumption that time- and space-like TFF coincide at $112$~GeV$^2$. The error we had to artificially add to reach at that conclusion is around a $20\%$, which agrees with our previous statements and also with \cite{Agaev:2014wna}.
Of course, adding this $20\%$ error in Eq.~(\ref{babarmixing}) solve what we call \babar \  puzzle.

A $15\%$ departure from the asymptotic limit may seem too large for that high momentum transfer. Notice~\cite{Agaev:2014wna,Bakulev:2000uh} that due to its nature, TFF are a convolution of a a perturbative hard-scattering amplitude and a gauge-invariant meson distribution amplitude (DA)~\cite{Mueller:1994cn} which incorporates the nonperturbative dynamics of the QCD bound-state~\cite{Lepage:1980fj}. That means that even for large $Q^2$ well inside the asymptotic region, soft scales coming from the Fock decomposition can enhance the TFF. These soft corrections depend on the broadness of the DA. At low energies, our fits suggest the typical hadronic scale for the $\eta$ TFF to be lower than the $\eta'$ counterpart. Being the $\eta'$ more contaminated by $s\bar{s}$ content (and less from other Fock states), one would expect its hadronic scale to be close to the $\phi$ meson mass, around $1$~GeV. This is in fact what we find, and indicates a narrower DA for the $\eta'$, dominated by a $q\bar{q}$ state, explaining at once why the duality arguments hold better than in the $\eta$ case. This argument complements the one discussed in  \cite{Agaev:2014wna} from the perturbative study of the TFFs.

Even larger error should be added to duality arguments at lower energies, such as the measurement of the CLEO Collaboration of the same cross section but at $\sqrt{s}=3.773$ GeV, and forthcoming measurements by the BES-III Collaboration at $\sqrt{s}=4.26$~GeV.

For all these reasons, we chose not to use these \babar \  measurements in the time-like region in our fits.

\subsection[A prediction for the $VP\gamma$ couplings]{A prediction for the {\boldmath $VP\gamma$} couplings}

In this subsection, we extend our analysis to the vector - pseudoscalar electromagnetic form factors. In particular, we are interested in the couplings of the radiative decays of lowest-lying vector mesons into $\eta$ or $\eta'$, i.e., $V \to (\eta,\eta')\gamma$, and of the radiative decays $\eta' \to V\gamma$, with $V=\rho,\omega,\phi$.

We follow closely the method presented in Refs.~\cite{Ball:1995zv,Escribano:2005qq}, and make use of the equations in Appendix A in Ref.~\cite{Escribano:2005qq} to relate the form factors with the mixing angle and the decay constants in the flavor basis. To account for the $\phi-\omega$ mixing we use $\phi_V=3.4^\circ$. The form factors, saturated with the lowest-lying resonance and then assuming vector meson dominance, can be expressed by
\begin{equation}
\label{gVPg}
F_{VP\gamma}(0,0)=\frac{f_V}{m_V}g_{VP\gamma}\, ,
\end{equation}
where $g_{VP\gamma}$ are the couplings we are interested in, and $f_V$ are the leptonic decay constants of the vector mesons and are determined from the experimental decay rates via
\begin{equation}
\Gamma(V\to e^+e^-)=\frac{4\pi}{3}\alpha^2 \frac{f_V^2}{m_V}c_V^2\, ,
\end{equation}
with $c_V$ an electric charge factor of the quarks that make up the vector, 
$c_V=(\frac{1}{\sqrt{2}}, \frac{\sin\theta_V}{\sqrt{6}},\frac{\cos\theta_V}{\sqrt{6}})$
for $V=\rho,\omega,\phi$ respectively.
Here $\theta_V = \phi_V+\arctan(1/\sqrt{2})$. Experimentally we find
\begin{equation}
\begin{split}
f_{\rho^0}=(221.2\pm0.9)\mathrm{MeV}\, ,\\[1ex]
f_{\omega}=(179.9\pm3.1)\mathrm{MeV}\, ,\\[1ex]
f_{\phi}=(239.0\pm3.8)\mathrm{MeV}\, .
\end{split}
\end{equation}
using $\Gamma(\rho \to e^+e^-) = 7.04(6)$~keV, $\Gamma(\omega \to e^+e^-) = 0.60(2)$ keV, and $\Gamma(\phi \to e^+e^-) = 1.27(4)$~keV from~\cite{Agashe:2014kda}.

The couplings in this flavor basis are:
\begin{equation}
\label{gcoupling}
\begin{split}
g_{\rho\eta\gamma}=\frac{3m_{\rho}}{4\pi^2f_{\rho^0}}\frac{\cos\phi}{\sqrt{2}F_q}\, , 
\qquad
g_{\rho\eta'\gamma}=\frac{3m_{\rho}}{4\pi^2f_{\rho^0}}\frac{\sin\phi}{\sqrt{2}F_q}\, ,\\[1ex]
g_{\omega\eta\gamma}=\frac{m_{\omega}}{4\pi^2f_{\omega}}\left(\cos\phi_V \frac{\cos\phi}{\sqrt{2}F_q} - 2 \sin\phi_V \frac{\sin\phi}{\sqrt{2}F_s} \right)\, ,\\[1ex]
g_{\omega\eta'\gamma}=\frac{m_{\omega}}{4\pi^2f_{\omega}}\left(\cos\phi_V \frac{\sin\phi}{\sqrt{2}F_q} + 2 \sin\phi_V \frac{\cos\phi}{\sqrt{2}F_s} \right)\, ,\\[1ex]
g_{\phi\eta\gamma}=-\frac{m_{\phi}}{4\pi^2f_{\phi}}\left(\sin\phi_V \frac{\cos\phi}{\sqrt{2}F_q} + 2 \cos\phi_V \frac{\sin\phi}{\sqrt{2}F_s} \right)\, ,\\[2ex]
\quad g_{\phi\eta'\gamma}=-\frac{m_{\phi}}{4\pi^2f_{\phi}}\left(\sin\phi_V \frac{\sin\phi}{\sqrt{2}F_q} - 2 \cos\phi_V \frac{\cos\phi}{\sqrt{2}F_s} \right)\, .\\[1ex]
\end{split}
\end{equation}
where we have assumed $\phi_q=\phi_s=\phi$. Table~\ref{tab:gVP} collects our predictions in its second column. Corrections due to $\phi_q \neq \phi_s$ to these formulae can be found in Appendix A, Eq. (A.5) of Ref.~\cite{Escribano:2005qq}. 

The decay widths of $P\to V\gamma$ and $V\to P\gamma$ are
\begin{equation}
\begin{split}
\Gamma(P\to V\gamma)&=\frac{\alpha}{8} g^2_{VP\gamma}\left(\frac{m_P^2-m_V^2}{m_P}\right)^3\, ,\\
\Gamma(V\to P\gamma)&=\frac{\alpha}{24} g^2_{VP\gamma}\left(\frac{m_V^2-m_P^2}{m_V}\right)^3\, .
\end{split}
\end{equation}

The experimental decay widths from \cite{Agashe:2014kda} allow us to extract an experimental value for $g_{VP\gamma}$, which are collected in the last column on Table~\ref{tab:gVP}.  

Our predictions compare  well with the experimental determinations, see Table~\ref{tab:gVP}, specially considering the simplicity of the approach. The differences are always below 2 standard deviations, excepting the $\omega$ couplings. Our prediction for the  ratio of $J/\Psi$ decays is in that respect remarkable. 

The observed deviations hint towards a somehow oversimplified approach. Even though our goal is just to show the relevance of TFF in other decays, and we do not pretend an exhaustive study of higher-order contributions in our scheme, we still want to remark two possible ways to improve our approach.

On the one hand, the fact that $F_q$ departs from $F_{\pi}$ in Eq.~\eqref{eq:mixingetap} may imply a correction through an OZI-violating parameter $\Lambda_1$ that appears at next-to-leading order in the Lagrangian of $\chi$PT Large-$N_c$ used to define the mixing equations, $F_q=F_{\pi}(1+\Lambda_1/3)$~\cite{Feldmann:1998sh,Feldmann:1999uf} which in turns imply $\phi_q \neq \phi_s$, since $\phi_q-\phi_s \sim \Lambda_1/3$. With the result in Eq.~\eqref{eq:mixingetap}, we estimate $\Lambda_1 \sim 0.2$, in agreement with the naive $1/N_c$ counting (i.e, $\Lambda_1 \sim 1/N_c \sim 0.3$), and then $\phi_q-\phi_s \sim 3.8^\circ$.

The ratio $R_{J/\Psi}$ provides direct information on the angle $\phi_q$ since
\begin{equation}
R_{J/\Psi}=\tan^2(\phi_q)\left(\frac{m_{\eta'}}{m_{\eta}}\right)^4\left(\frac{M^2_{J/\Psi}- m_{\eta'}^2 }{M^2_{J/\Psi}- m_{\eta}^2 } \right)^3\ ,
\end{equation}
with $M_{J/\Psi}$ the $J/\Psi$ meson mass. The experimental $R_{J/\Psi}$ ratio defined in last the row in Tab.~\ref{tab:gVP}, results in $\phi_q=(38.1\pm0.6)^{\circ}$, which implies $\phi_s=(38.1+3.8 \pm 1.6)^\circ = (41.9\pm1.6)^\circ$ with the error coming from our determination of $\phi$ in Eq.~\eqref{eq:mixingetap}. Even though both angles are distinguishable, their impact on the $g_{VP\gamma}$ is a shift of the form $g_{VP\gamma} \to g_{VP\gamma} /(\cos \phi_q \cos \phi_s + \sin \phi_q \sin \phi_s)$~\cite{Escribano:2005qq}. For the $\phi_q = \phi_s$ limit, such shift is exactly $1$. Using the $3.8^\circ$ difference, such shift translates into $0.998$, a $2$ per mil effect, negligible. Our assumption  $\phi_q = \phi_s=\phi$ is supported phenomenologically.

On the other hand, as discussed in detail in Ref.~\cite{Kaiser:2000gs}, in the flavor singlet channel one has to allow for another OZI-rule violating correction, which essentially corresponds to replacing $F_0 \to F_0/(1 + \Lambda_3)$. This shifts both the $P\rightarrow\gamma\gamma$ decays and the formulae for $g_{VP\gamma}$ predictions~\cite{Feldmann:1999uf}. The parameter $\Lambda_3$ is, however, still unknown, although expected to be $\sim 1/N_c \sim 0.3$. We can make use of Eq.~\eqref{eq:degenRUN} to estimate it. The shift on $F_0$ can be translated into a shift in $F_{q,s}$ recalling that both are related to $F_{\pi}$ following Eqs.~(\ref{08results},\ref{eq:mixresults}) respectively, and find $F_{q,s}\to F_{q,s}/(1 \pm \Lambda_3)$ as well.  Going then to Eq.~\eqref{eq:FLAVEtaDecay}, $F_{\eta(')\gamma\gamma}(0) \to F_{\eta(')\gamma\gamma}(0) (1-\Lambda_3)$.
 
Then, Eq.~\eqref{eq:degenRUN} transforms into 
\begin{equation}
\label{eq:degenRUNL3}
\begin{split}
\lim_{Q^2\to\infty }Q^2\left( F_{\eta\gamma^*\gamma}(Q^2)F_{\eta\gamma\gamma}(0) +  F_{\eta'\gamma^*\gamma}(Q^2) F_{\eta'\gamma\gamma}(0) \right) \\ \times(1-\Lambda_3)= \left(1+\frac{8}{9}\delta \right)\frac{3}{2\pi^2} \, ,
\end{split}
\end{equation}
which after expanding and reorganizing in such a way that in the l.h.s. remains only experimental quantities, results in:
\begin{equation}
\label{deltaL3exp}
\begin{split}
\lim_{Q^2\to\infty }Q^2\left( F_{\eta\gamma^*\gamma}(Q^2)F_{\eta\gamma\gamma}(0) +  F_{\eta'\gamma^*\gamma}(Q^2) F_{\eta'\gamma\gamma}(0) \right) \\ = \left(1+\frac{8}{9}\delta + \Lambda_3 +\frac{8}{9}\delta \Lambda_3 \right)\frac{3}{2\pi^2} \, .
\end{split}
\end{equation}

We recall that l.h.s., experimentally, reads $0.89 \frac{3}{2\pi^2} $, and $\delta = -0.17$. With \eqref{deltaL3exp} we find $\Lambda_3=0.05$, smaller than expected and with positive sign.

The VP$\gamma$ couplings are also shifted by $\Lambda_3$. The expressions can be found in Eq.~(42) in Ref.~\cite{Feldmann:1999uf} which, after expanding, can be expressed as a shift on the couplings in our Eq.~\eqref{gcoupling}: $g_{V\eta\gamma} \rightarrow (g_{V\eta\gamma} + |g_{V\eta\gamma}|\Lambda_3/2)$ and $g_{V\eta'\gamma} \rightarrow  (g_{V\eta'\gamma} + |g_{V\eta'\gamma}|\Lambda_3)$, always increasing the coupling.  For some of them, the $\Lambda_3$ correction goes on the right direction (the $\rho$ case), but for others it is not conclusive (the $\phi$ case where for $\eta$ goes well and for $\eta'$ wrong). The result of the shift is, then, ambiguous.

Discarding OZI-violating effects, Pad\'e approximants can then be the avenue to follow since the vector mass that should be used in Eq.\eqref{gVPg} it should not correspond to a physical observable, but an effective scale provided by the pole of a PA assuming the philosophy of the present work. For the $\eta$ TFF, the $\Lambda_{\eta}^2$ from~\eqref{eq:tffpole} is smaller than the vector meson dominance mediator. If the same would happen with the $\rho,\omega$ form factors, one would expect, then, different $g_{VP\gamma}$ couplings. Since this study is beyond the scope of the present analysis, we postponed for future work. A naive estimate of these effects could be accounted for within the half-width-rule~\cite{Masjuan:2012sk}, i.e., instead of using $m_V$ in~\eqref{gVPg}, use $m_V \pm \Gamma_V/2$, with $\Gamma$ the full width of the vector. This provides a way to asses the error of neglecting the width of the resonance in using $m_V$. For example, for the $\rho$ case, within the half-width-rule, the errors of the $g_{\rho P\gamma}$ would be enlarged by a factor 3, well compatible with the experimental determinations.

Further studies along these lines are postponed for future work.
\begin{table}[h]
\centering
\label{tab1}       
\begin{tabular}{ccc}
& Prediction & Experiment\\ [3pt]
\hline\\ [-5pt]
$g_{\rho \eta \gamma}$ & $1.50(4)$ & $1.58(5) $ \\
$g_{\rho \eta^\prime \gamma}$ & $1.18(5)$ & $1.32(3) $ \\
$g_{\omega \eta \gamma}$ & $0.57(2)$ & $0.45(2) $ \\
$g_{\omega \eta^\prime \gamma}$ & $0.55(2)$ & $0.43(2) $ \\
$g_{\phi \eta \gamma}$ & $-0.83(11)$ & $-0.69(1) $ \\
$g_{\phi \eta^\prime \gamma}$ & $0.98(14)$ & $0.72(1) $ \\ [3pt]
\hline\\[-5pt]
$R_{J/\Psi}=\frac{\Gamma(J/\Psi \to \eta^\prime \gamma)}{\Gamma(J/\Psi \to \eta \gamma)} $ & $4.74(55)$ & $4.67(20)$\\ [5pt]
\hline
\end{tabular}
\caption{Summary of VP$\gamma$ couplings. Experimental determinations are from Ref.~\cite{Agashe:2014kda}.}
\label{tab:gVP}
\end{table}

\section{Conclusions}
\label{sec:conc}
In the present work, the $\eta$ transition form factor has been analyzed for the first time in both space- and time-like
regions at low and intermediate energies making use of a model-independent approach based on the use of
rational approximants of Padé type.
The model independence of our approach is achieved trough a detailed and conservative evaluation of the
systematic error associated to it.
The new set of experimental data on the $\eta\to e^+e^-\gamma$ reaction provided by the A2 Collaboration
in the very low-energy part of the time-like region allows for a much better determination of the slope and curvature 
parameters of the form factor, as compared to the predictions obtained in our previous work only using space-like data,
which constitute the most precise values up-to-date of these low-energy parameters.
Our method is also able to predict for the first time the third derivative of the form factor.
In addition, the new analysis has served to further constrain its values at zero momentum transfer and infinity.
We have seen that our results, in particular for the case of the slope parameter, are quite insensitive to
the values used in the fits for the two-photon decay width of the $\eta$, thus showing that the collection of
space- and time-like experimental data is more than enough to fix a value for the normalization of the form factor 
compatible with current measurements.
We have also seen that the role played by the high-energy space-like data is crucial to get accurate predictions
for the low-energy parameters of the form factor and its asymptotic value.
As a consequence of these new results, we have fully reanalyzed the $\eta$-$\eta^\prime$ mixing parameters
this time also considering renormalisation-scale dependent effects of the singlet decay constant $F_0$.
The new values obtained are already competitive with standard results having the advantage of requiring much less
input information.
Related to this, we have also obtained predictions for the $VP\gamma$ couplings which are in the ballpark of
present-day determinations.

In summary, the method of Padé approximants has been shown to be very powerful for fixing the low-energy
properties of the $\eta$ transition form factor making their predictions more accurate and well-established.
This fact opens the door to a more exhaustive analysis of the single Dalitz decay processes
$P\to l^+l^-\gamma$, with $P=\pi^0, \eta, \eta^\prime$ and $l=e, \mu$, the double Dalitz ones
$P\to l^+l^-l^+l^-$ (in all possible kinematically allowed configurations) \cite{RE&SGS},
and the rare lepton-pair decays $P\to l^+ l^-$ ---see the $\pi^0\to e^+ e^-$ application in Ref.~\cite{Masjuan:2015lca}, 
which are usually discussed only in terms of monopole approximations.
Indeed, when this work was being concluded the BESIII Collaboration reported a first observation of the
$\eta^\prime\to e^+e^-\gamma$ process measuring the branching ratio and extracting the
$\eta^\prime$ transition form factor~\cite{Ablikim:2015wnx}.
This new measurement may put our approach with its back to the wall.
However, a very preliminary analysis of this recent data in comparison with our prediction for this form factor
in the time-like region exhibits a nice agreement but reveals the necessity of going beyond the
vector meson dominance model used in the experimental analysis~\cite{RE&SGS&PM&PSP}.

\appendix

\section{Best Pad\'e approximant fit parameterisation}
\label{AppTL}

In this appendix we provide the parameterizations of our best $P^L_1(Q^2)$ fit for the $Q^2 F_{\eta\gamma^*\gamma}(Q^2)$. Defining $P^L_1(Q^2)$ as:  
\begin{equation}\label{PL1}
P^L_1(Q^2)\, =\, \frac{T_N(Q^2)}{R_1(Q^2)}\, =\, \frac{t_1 Q^2+ t_2 Q^4+\cdots t_N (Q^2)^N}{1+r_1 Q^2}\, ,
\end{equation}
the corresponding fitted coefficients\footnote{For full precision of the coefficients together with the correlation matrix, contact the corresponding authors.} for  the $Q^2 F_{\eta\gamma^*\gamma}(Q^2)$ are collected in Table~\ref{tab:param}.

\begin{table}[ht]
\centering
\begin{tabular}{ | c | c |}
\hline
 &     $\eta$-TFF                \\ \hline
        \hspace{0.2cm}                                     $t_1$                   \hspace{0.2cm}                    &  \hspace{0.3cm}    $0.27349$   \hspace{0.3cm}\\ 
                                         $t_2$                                   & $1.1771\cdot 10^{-2}$  \\
                                        $t_3$                                         & $-1.1048\cdot 10^{-3}$   \\
                                        $t_4$				  & $2.8861\cdot 10^{-5}$  \\
                                        $t_5$				   & $2.2974\cdot 10^{-6}$  \\
                                        $t_6$				   &  $-1.5096\cdot 10^{-7}$  \\ 
                                        $t_7$				   &  $2.3655\cdot 10^{-9}$  \\  
                                         \hline
                                      $r_1$				   & $1.9584$   \\
\hline                                           
\end{tabular}
\caption{Fitted coefficients for our best $P^7_1(Q^2)$ for the $Q^2 F_{\eta\gamma^*\gamma}(Q^2)$.}
\label{tab:param}
\end{table}

With the coefficients in Table~\ref{tab:param}, one can extract the slope of the TFF by expanding~(\ref{PL1}) and normalizing the result as in Eq.~(\ref{eq:tffexpansion}):
\begin{equation}
b_{\eta} = (t_1 \cdot r_1 - t_2) m_{\eta}^2/t_1 = 0.5749
\end{equation}
with $m_{\eta} = 0.547853$~GeV, to be compared with the third column in Table~\ref{tab:psresults}.

\section{Convergence of the Pad\'e approximant sequence}
\label{AppConv}
To test how fast the convergence of our PA sequence is we analyze here a simple holographic confining model presented in~\cite{Brodsky:2011xx} (and also explored in Ref.~\cite{Brodsky:2011yv}), based on light-front holographic QCD where the correct small $Q^2$ behavior (in order to simulate confinement) is introduced using the dressed current (see~\cite{Brodsky:2011xx} for details)\footnote{We do not consider higher-twist components here to keep the model easy to use.}.

In this context, the TFF is defined as (assuming, for simplicity, $\eta \sim \eta_8$)
\begin{equation}
\label{LFmodel}
F_{ \eta \gamma^*\gamma}(Q^2)=\frac{P_{q\bar{q}}}{\sqrt{3} \pi^2 f_{\pi}} \int_0^1 \frac{dx}{(1+x)^2} x^{Q^2 P_{q\bar{q}}/(8\pi^2 f_{\pi}^2)}\, ,
\end{equation}
where $P_{q\bar{q}}$ is the probability of finding the $q\bar{q}$ component in the $\eta$ light-front wave function, and we impose $P_{q\bar{q}}=0.5$ for numerics.

\begin{figure*}[tbp]
\begin{center}
	\includegraphics[width=0.45\textwidth]{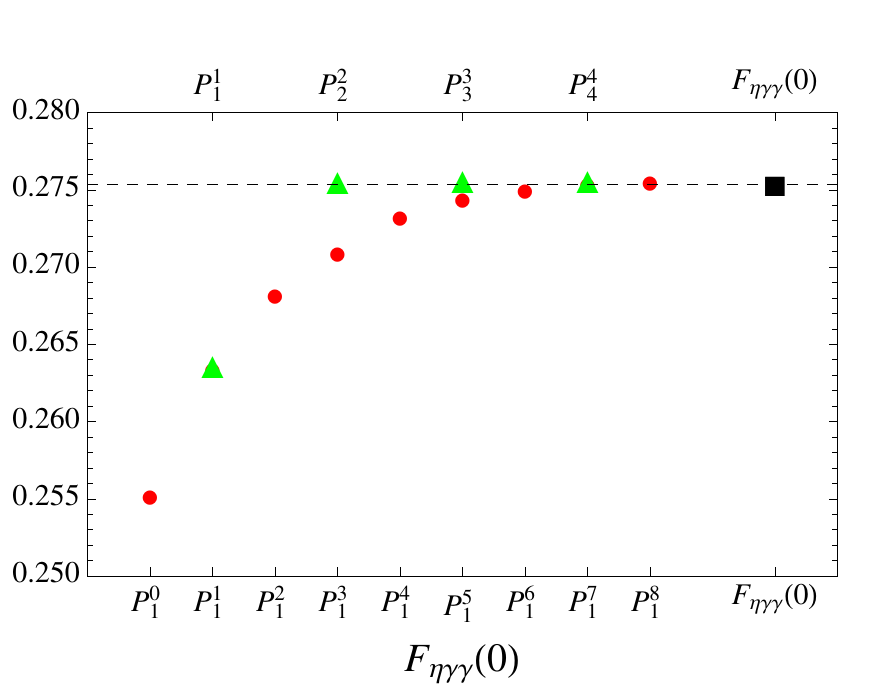}
	\includegraphics[width=0.45\textwidth]{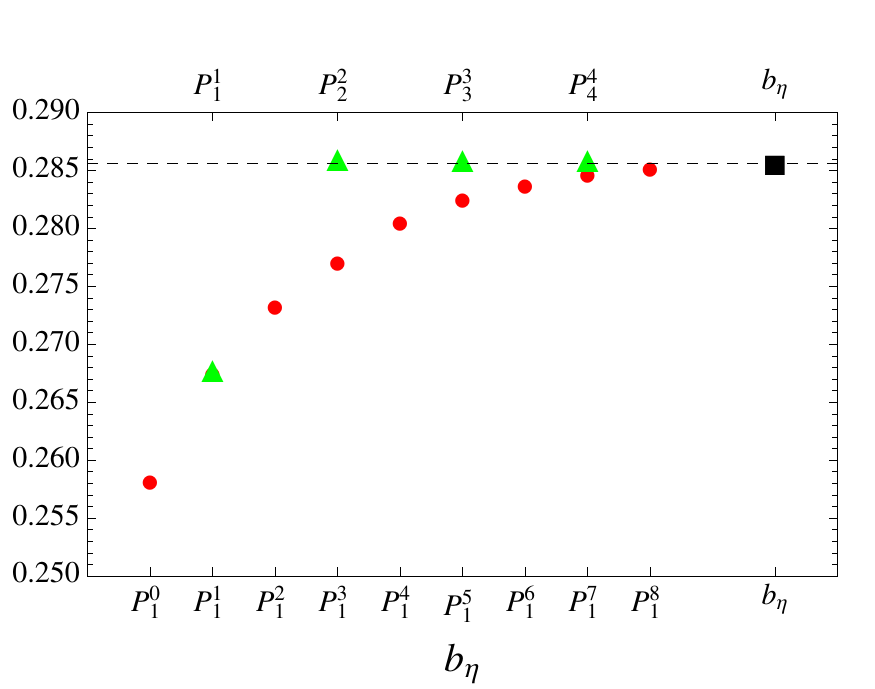}\\
	\includegraphics[width=0.45\textwidth]{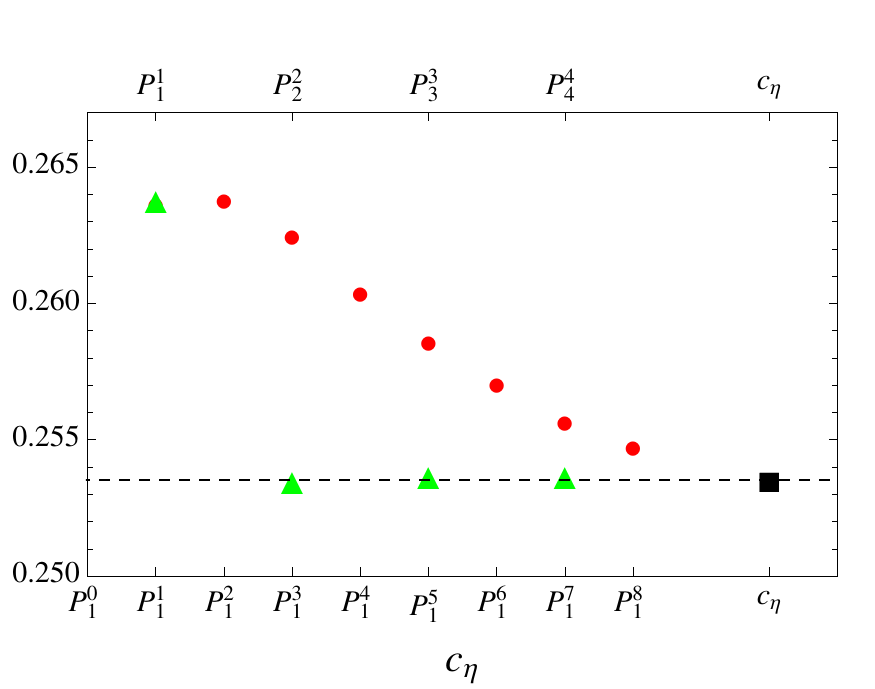}
	\includegraphics[width=0.45\textwidth]{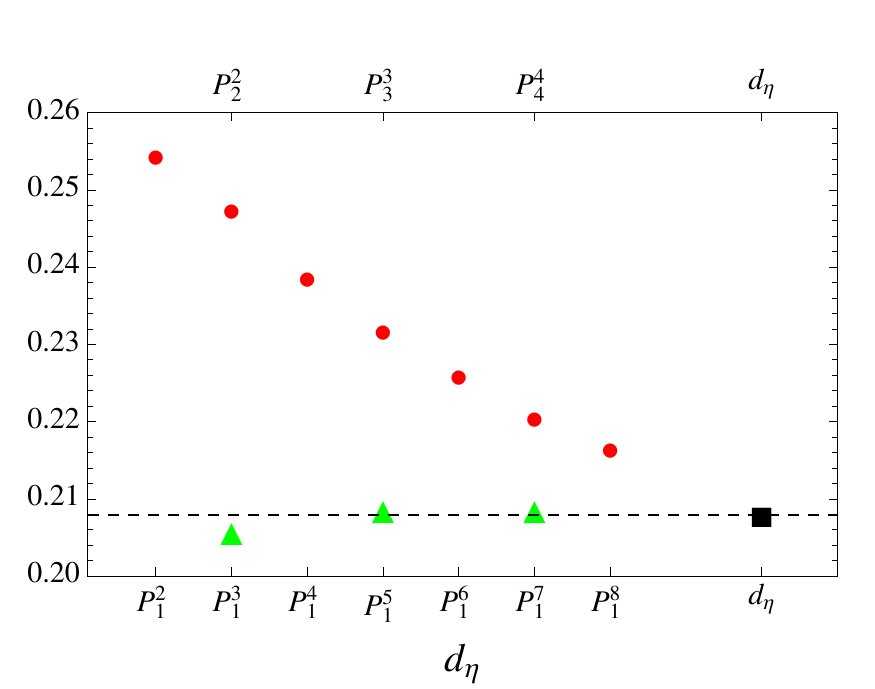}
\caption{Convergence pattern for the $P^L_1$ (red circles) and $P^N_N$ (green triangles) sequences to $F_{\eta \gamma \gamma}(Q^2)$ for $Q^2=0$ and its firsts three derivatives $b_{\eta},c_{\eta}$ and $d_{\eta}$. The black squares show the results for each parameter from the model in Eq.~(\ref{LFmodel}), prolonged by the dashed line.}
\label{fig:convergence}
\end{center}
\end{figure*}

Once the model is defined, by generating a set of pseudodata as it is done is Sec.~\ref{sec:syserror} we can test how fast the PA sequence converge to $Q^2 F_{\eta\gamma^*\gamma}(Q^2)$. We fit these pseudodata with both $P^L_1(Q^2)$ and $P^N_N(Q^2)$ sequences going up to $L=8$ and $N=4$ and we show the convergence pattern for the value of the $\eta$ TFF at the origin, $F_{\eta \gamma \gamma}(0)$, and its three first derivatives ($b_{\eta}, c_{\eta}$, and $d_{\eta}$) in Fig.~\ref{fig:convergence} (the black square shows the corresponding value from Eq.~(\ref{LFmodel})). This exercises complements the one studied in the Appendix of Ref.~\cite{Escribano:2013kba}. The first element of the sequence, the $P^1_1(Q^2)$ contains only two parameters and then only $F_{\eta \gamma \gamma}(0)$ and $b_{\eta}$ can be directly extracted from the fit. By reexpanding it one could predict all the other derivatives but we only show fit outputs. The same applies for the second element of the sequence, the $P^1_1(Q^2)$ and its third derivative. The reader should notice the hierarchy pattern of convergence. While $F_{\eta \gamma \gamma}(0)$ is approach very fast and with the $P^2_1(Q^2)$ the error is about a $2\%$, the derivatives are worse predict, being the third one the worst. Comparing each prediction with its counterpart from the model gives and idea of the systematic error done by such procedure (see Ref.~\cite{Escribano:2013kba} for details about the systematic error from our method). The $P^N_N(Q^2)$ sequence converges much faster, reaching the $10^{-3}$ with its second element. The inconvenience is its growing two-by-two inputs for each new element, making the sequence more difficult to be performed in fits to real data. These results are model dependent and other models may yield different systematic errors. The trend shown in Fig.~\ref{fig:convergence} is, however, general although models such us the logarithmic one studied in Ref.~\cite{Masjuan:2012wy} converge faster (reaching smaller errors) even though the systematic error for the first elements on the $P^L_1(Q^2)$ are larger.

\begin{acknowledgements}
We thank Susan Gardner, Santi Peris and Marc Unverzagt for discussions.
This work was supported in part by the Deutsche Forschungsgemeinschaft DFG 
through the Collaborative Research Center ``The Low-Energy Frontier of the Standard Model'' (SFB 1044).
This work was also supported in part by 
the Ministerio de Ciencia e Innovaci\'on under grant FPA2011-25948,
the Secretaria d'Universitats i Recerca del Departament d'Economia i
Coneixement de la Generalitat de Catalunya under grant 2014 SGR 1450,
the Ministerio de Econom\'{\i}a y Competitividad under grant SEV-2012-0234,
the Spanish Consolider-Ingenio 2010 Programme CPAN (CSD2007-00042), and
the European Commission under programme FP7-INFRASTRUCTURES-2011-1
(Grant Agreement N. 283286).
\end{acknowledgements}


\end{document}